\documentclass[a4paper,11pt]{article}
\usepackage{jheppub} %
\usepackage{amsmath,amssymb,amsfonts,stmaryrd}
\usepackage[dvipsnames]{xcolor}
\usepackage[normalem]{ulem}

\preprint{\texttt{KIAS-P25044}}

\title{\boldmath Classical eikonal in relativistic scattering}

\author[a]{Sungsoo Kim,}
\author[b]{Hojin Lee,}
\author[b]{Sangmin Lee}
\affiliation[a]{Department of Physics and Astronomy, Seoul National University, \\
1 Gwanak-ro, Gwanak-gu, Seoul 08826, Korea}
\affiliation[b]{School of Physics, Korea Institute for Advanced Study, \\
85 Hoegiro, Dongdaemun-Gu, Seoul 02455, Korea}

\emailAdd{sooo4017@snu.ac.kr, hojinlee@kias.re.kr, sangminlee@kias.re.kr}

\abstract{The classical eikonal is defined to be the generator of all scattering observables in a scattering problem in classical mechanics.
It was originally introduced as the log of the quantum S-matrix in the classical limit. 
But its classical nature calls for a definition and computational methods independent of quantum mechanics. 
In this paper, we formulate a classical interaction picture which serves as the foundation of the classical eikonal. Our emphasis is on generality. 
In perturbation theories, both Hamiltonian deformation and symplectic deformation are considered. 
Particles and fields are treated on a similar footing. The causality prescription of the propagator is essentially the same for non-relativistic and relativistic kinematics. For a probe particle in electromagnetic or gravitational background, we present all order formulas for the perturbative eikonal. In the electromagnetic setting, 
we also illustrate how the eikonal encodes the information on radiation of external fields.}

\begin{document}
\maketitle
\flushbottom

\section{Introduction and outlook}
\label{sec:intro}

For the past decade or so, a variety of tools for scattering amplitudes have been applied to classical general relativity (GR) to predict gravitational wave signals from binary systems;
see {\it e.g.} \cite{Bjerrum-Bohr:2022blt,Kosower:2022yvp,Buonanno:2022pgc} for reviews. 
Scattering amplitudes are inherently quantum, but for many applications in GR, we only seek observables in the classical limit. It is only natural to reformulate classical mechanics (including GR) 
in ways inspired by scattering amplitudes without actually having to take the classical limit of a quantum theory. 

A particular classical avatar of quantum scattering amplitudes was put forward in \cite{Gonzo:2024zxo, Kim:2024grz,Kim:2024svw,Alessio:2025flu} and dubbed ``classical eikonal" in \cite{Kim:2024grz,Kim:2024svw}. 
In any classical scattering problem in Hamiltonian formulation, the classical eikonal is defined as the generator of canonical transformation from in-states to out-states. By states we mean points in the asymptotic phase space of \emph{free} particles or fields. 
The classical eikonal was originally motivated by eikonal approximation and exponential representation of quantum S-matrix \cite{Feynman:1951gn,Lehmann:1957zz,Damgaard:2021ipf,Damgaard:2023ttc}, but it can certainly be defined within classical mechanics; it could have been introduced in the 19th century.

When a binary system allows for a separable Hamilton-Jacobi action, the \emph{radial action} coincides with the classical eikonal. 
But, the notion of classical eikonal is far more general and unrestricted by the number of particles/fields or the symmetries of the system under study. 
In \cite{Kim:2024svw}, it was shown that the Magnus expansion \cite{Magnus:1954zz,BLANES2009151,ebrahimifard2023magnusexpansion} is a powerful tool 
to compute the classical eikonal perturbatively. 
The diagrammatic rule of the Magnus expansion is quite close to those arising from worldline quantum field theory (WQFT) 
\cite{Mogull:2020sak}, but the causality prescription of propagators turns out to be subtle and 
introduces non-rooted as well as rooted trees with specific weights. 
This contrasts with the (Berends-Giele) diagrammatic expansion of the equation of motion, 
where all causality arrows flow toward the final observable, forming only rooted trees \cite{Jakobsen:2022psy}. 

In \cite{Kim:2024svw}, the structure of the Magnus expansion was elucidated based on a Newtonian scattering problem as a toy model.
The structure was then argued to be equally applicable to a relativistic binary problem. 
The third order post-Minkowskian (3PM) eikonal was computed as an example, with nontrivial relative weights among different causal diagrams, and shown to agree with independent computations in the literature. 
Much like WQFT, the classical eikonal treats particles and fields on a similar footing, 
hence it can be generalized to include radiation observables as shown in \cite{Kim:2025hpn}. 

The goal of this paper is to elaborate further on aspects of the classical eikonal and methods to compute it. 
The main difference compared to \cite{Kim:2024svw} is the relativistic kinematics 
and the inclusion of fields from the lowest order of perturbation. 
In section~\ref{sec:cip}, we set out to define the eikonal in classical mechanics with minimal assumptions. Then we set up the ``classical interaction picture", a classical analog of the interaction picture familiar from quantum mechanics. 
When a solvable model is subject to a smooth deformation, a vector field on the phase space can be defined, which by an exponential map generates the flow from the unperturbed trajectory to the perturbed one. 

When the Hamiltonian is deformed for a fixed symplectic form, the map is always a canonical transformation, admitting the classical eikonal even for a finite-time evolution. It is equally well defined for bound and unbound orbits.
In contrast, when the symplectic form is deformed for a fixed Hamiltonian, to keep manifest gauge invariance of equation of motion, the map between the two trajectories may not define a canonical transformation. In a scattering problem where the deformation dies out asymptotically, the map converges to a canonical transformation which allows us to ``extract" the classical eikonal. 

In section~\ref{sec:EM-background} (electromagnetic field background in flat spacetime) and \ref{sec:Grav-background} (weakly curved background spacetime), we discuss the classical eikonal of a relativistic probe particle. 
Compared to the Newtonian case, one notable difference is that we have to deal with constraints; even a point particle in special relativity is constrained. The constraints can be treated using Dirac brackets as illustrated in \cite{Gonzo:2024zxo,Kim:2024grz,Alessio:2025flu}. When the constraints form pairs of ``gauge generators" and ``gauge-fixing functions", %
the difference between the Poisson bracket and the Dirac bracket vanishes for gauge-invariant observables. 

The electromagnetic background in section~\ref{sec:EM-background} serves as a typical example 
of symplectic deformation. During the time evolution, the vector field (on phase space) for the classical interaction picture is well defined but is not generated by a (gauge-invariant) potential. 
When we perform the Magnus expansion, we have to work with Lie brackets of the vector fields. 
Only in the end, we can ``extract" the classical eikonal for the scattering process. 
Aside from this issue, the final result of the Magnus expansion agrees with the Feynman rules of WQFT 
augmented by the causality prescription of \cite{Kim:2024svw}.

In section~\ref{sec:Grav-background}, we study the probe motion in a weakly curved spacetime. The problem can be regarded as a Hamiltonian deformation, so the Magnus expansion can be applied straightforwardly. The comparison with the WQFT is slightly complicated by the fact that the natural metric perturbation is $g^{\mu\nu} = \eta^{\mu\nu} - \tilde{h}^{\mu\nu}$ for the Magnus expansion, unlike $g_{\mu\nu} = \eta_{\mu\nu} + h_{\mu\nu}$ of the WQFT. 
Of course, the two approaches are based on the same action functional, so they must agree on the final result for the classical eikonal. However, in intermediate steps, the Magnus approach appears to carry fewer diagrams.

In section~\ref{sec:field}, in electromagnetism for simplicity, we explain how radiative processes can also be described by the classical eikonal. In \cite{Kim:2025hpn}, in the context of gravitational binary, the leading (1.5PM) radiation eikonal was obtained and shown to account for the radiated momentum and the radiation loss of the particle momenta. 
Our aim is not to compute a new observable at a higher order, but to demonstrate the fact that 
\emph{all} radiative processes can be described by the classical eikonal. 
We show a few simple examples including the Compton scattering. 

The classical eikonal is applicable without limits on the number of degrees of freedom or the symmetry of the problem. However, symmetries put restrictions on the form of the eikonal and make it easier to compute. 
Recent works \cite{Witzany:2019nml,Compere:2021kjz,Compere:2023alp,Akpinar:2025tct} on the integrability of spinning black hole binary look particularly promising and deserve further study. 
One could also use the classical eikonal to analyze bound orbits arising naturally from post-Newtonian (PN) expansion. We hope to address these open directions in future work.

\section{Classical interaction picture} \label{sec:cip}

In \cite{Kim:2024grz,Kim:2024svw}, 
the classical eikonal was defined as the classical limit 
of the log of the S-matrix:
\begin{align}
    \chi = \lim_{\hbar \rightarrow 0} \frac{\hbar}{i}\log \hat{S} \,.
\end{align}
Our goal here is to set up a classical analog of the \emph{interaction picture} familiar from quantum mechanics 
in order to define $\chi$ purely within classical mechanics. 
Compared to \cite{Kim:2024svw}, the discussion here is expanded in three directions. 
First, we consider an arbitrary Hamiltonian deformation $H=H_0 + H_1$, 
where $H_0$ does not necessarily describe a free particle. 
Second, we discuss not only the Hamiltonian deformation but also the symplectic deformation. Third, we elaborate on the fact that fields can be treated on 
an equal footing as particles.

Before we start developing the classical interaction picture (CIP), 
we briefly review some aspects of WQFT and 
discuss how CIP naturally arises from WQFT.

\subsection{WQFT Feynman rules} 

\paragraph{Setup} 
We consider a broad class of action functional in Hamiltonian formulation, 
\begin{align}
    S = S_\omega + S_H = \int \left[ \theta_m(\zeta) \dot{\zeta}^m - H(\zeta)  \right] dt \,.
\end{align}
Here, $\zeta^m$ are coordinates on a classical phase space, which combine all $x$ and $p$ coordinates. 
The simplest example is $(\zeta^1, \zeta^2) = (x,p)$ with $\omega^{12}= -\omega^{21} = 1$. 
The symplectic potential $\theta = \theta_m d\zeta^m$ is related to the symplectic form as usual, $\omega = d\theta$. 
Constraints are one of the reasons why we prefer Hamiltonian formulation. We do not discuss constraints explicitly here, but they can be incorporated using standard methods such as Dirac brackets. 

WQFT computes the observables in the quantum theory using path integral 
with a background-fluctuation expansion, 
\begin{align}
    \zeta(t) = \bar{\zeta}(t) + \delta\zeta(t) \,.
\end{align}
The fluctuation gives rise to vertex factors  
both from the symplectic potential and from the Hamiltonian. 
From the symplectic form, up to total derivatives, we find 
\begin{align}
    \begin{split}
    S_\omega &=\sum_{r=0}^\infty \frac{1}{r!} \int  \mathcal{V}_\omega^{(r)} \, dt \,,
\\
        \mathcal{V}^{(0)}_\omega &= \theta_m \dot{\bar{\zeta}}^m \,,
        \\
        \mathcal{V}^{(1)}_\omega &=  [\omega_{mn} \dot{\bar{\zeta}}^n] \delta \zeta^m  \,,
        \\ 
        \mathcal{V}^{(2)}_\omega &=  [\partial_{m_1} \omega_{m_2 n} \dot{\bar{\zeta}}^n] \delta \zeta^{m_1} \delta \zeta^{m_2}  +  \omega_{m_1 m_2} \delta \zeta^{m_1} \delta \dot{\zeta}^{m_2}  \,,
        \\
        &\;\, \vdots 
        \\
        \mathcal{V}^{(r)}_\omega &= [\partial_{m_1} \cdots \partial_{m_{r-1}} \omega_{m_r n}  \dot{\bar{\zeta}}^n] \,\delta \zeta^{m_1} \cdots \delta \zeta^{m_r}
        \\
        &\qquad + (r-1)\, [\partial_{m_1} \cdots \partial_{m_{r-2}} \omega_{m_{r-1} m_r} ]  \,\delta \zeta^{m_1} \cdots \delta \zeta^{m_{r-1}} \delta \dot{\zeta}^{m_r} \,.
    \end{split}
    \label{vertex-omega}
\end{align}
From the Hamiltonian, we find
\begin{align}
    \begin{split}
    S_H &= \sum_{r=0}^\infty \frac{1}{r!} \int  \mathcal{V}_H^{(r)} \, dt \,,
\\
        \mathcal{V}^{(0)}_H &= -H(\bar{\zeta}) \,,
        \\
        \mathcal{V}^{(1)}_H &= -(\partial_m H) \delta \zeta^m  \,,
        \\ 
        \mathcal{V}^{(2)}_H &= - (\partial_{m_1} \partial_{m_2} H )  \delta \zeta^{m_1} \delta \zeta^{m_2}
        \\
        &\;\, \vdots 
        \\
        \mathcal{V}^{(r)}_H &= - (\partial_{m_1} \cdots \partial_{m_r} H) \, \delta \zeta^{m_1} \cdots \delta \zeta^{m_r}\,.
    \end{split}
    \label{vertex-H}
\end{align}
One typically assumes that $\bar{\zeta}$ solves the EOM. Then the terms linear in $\delta \zeta$ all vanish. The quadratic terms define a propagator, and the cubic or higher terms define interaction vertices. 

The classical action often consists of a ``solvable" part and a ``deformed" part:
\begin{align}
    \omega = \omega^\circ + \omega' 
    \quad 
    \mbox{and/or} 
    \quad 
    H = H_0 + H_1 \,.
    \label{solvable-vs-deformed}
\end{align}
If one has to work with a background $\bar{\zeta}$ which is a solution to the solvable part only, 
the interaction vertices can have linear and quadratic terms coming from the deformed part. 

In many practical applications, both $\omega^\circ$ and $H_0$ are quadratic in coordinates, 
so that all interaction vertices come from the deformation. 
But, in principle, we are free to choose any solvable theory and add deformation. 
In general, we have to include cubic or higher vertices from  $\omega^\circ$ and $H_0$ 
in addition to the vertices from $\omega'$ or $H_1$. 

In the context of relativistic binary systems, 
examples of non-quadratic $(\omega^\circ, H_0)$ are common. 
In the post-Newtonian (PN) expansion, 
the solvable (Newtonian) Hamiltonian is certainly not quadratic:
\begin{align}
    H_0 = H_\mathrm{N} = \frac{\vec{p}^2}{2\mu} - \frac{GM\mu}{r} \,,
    \quad 
    H_1 = H_\mathrm{PN} = H_\mathrm{1PN} + H_\mathrm{2PN} +\cdots \,. 
    \label{PN-Hamiltonian}
\end{align}
In the post-Minkowskian (PM) expansion, 
if we focus on spinless particles, $(\omega^\circ, H_0)$ are effectively quadratic. 
But, if we consider spinning particles, $\omega^\circ$ is no longer quadratic \cite{Kim:2024grz}.

\paragraph{EOM vs. Eikonal} 

\begin{figure}[htbp]
    \centering
    \includegraphics[width=0.64\linewidth]{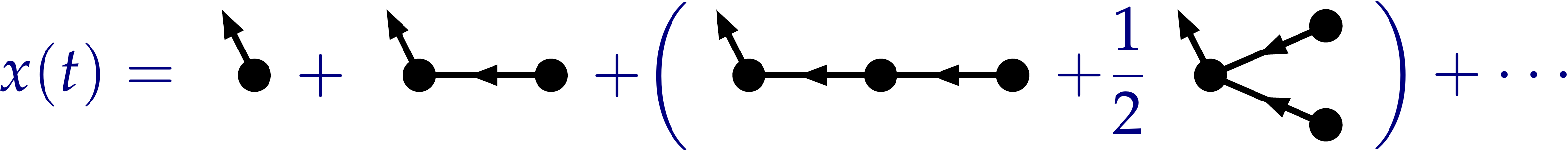}
    \caption{Rooted trees for a perturbative solution to the EOM.}
    \label{fig:EOM-rooted}
\end{figure}

WQFT offers a way to solve the classical EOM in terms of Feynman diagrams. 
The 1-point function of the interacting theory in the classical limit (sum over tree diagrams only) solves the EOM. 
In the Newtonian example, 
\begin{align}
    H_0 = \frac{p^2}{2m} \,,
    \quad 
    H_1 = V(x) \,,
\end{align}
a diagrammatic representation of the perturbative solution, 
\begin{align}
    x(t) = \bar{x}(t) + x_{(1)}(t) + x_{(2)}(t) + \cdots \,,
    \quad 
    \bar{x}(t) = \bar{x} + \bar{v} t \,,
\end{align}
is shown in Figure~\ref{fig:EOM-rooted}. 
The ``external legs" represent $x_{(n)}(t)$ which we wish to compute. 
The propagators are the retarded Green's function for $(d/dt)^2$ 
and 
the vertex factors are derivatives of $V(x)$ as in \eqref{vertex-H}. 
The coefficients are nothing but the Feynman symmetry factors. 
The Feynman rules do not specify the causality of the propagators. 
For an initial value problem in classical mechanics, the ``all thing retarded" prescription is suitable; all the arrows flow toward the final observable to be computed. From a QFT perspective, it is related to the Schwinger-Keldysh (in-in) formalism \cite{Jakobsen:2022psy}. 

In the same setting, the classical eikonal is the sum over all connected tree diagrams. 
The causality prescription turns out to be more subtle. 
As shown in \cite{Kim:2024svw}, the sum involves not only the rooted trees but also the non-rooted trees. 
The coefficients, known as Murua's $\omega$-coefficients \cite{Murua_2006}, are some refinements of the Feynman symmetry factors.

\begin{figure}[htbp]
    \centering
    \includegraphics[width=0.75\linewidth]{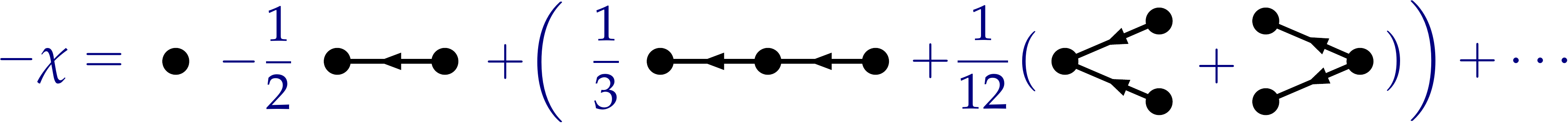}
    \caption{Non-rooted trees for a perturbative construction of the eikonal.}
    \label{fig:Eikonal-non-rooted}
\end{figure}

We should stress that this structure is quite universal. At least for quadratic $(\omega^\circ, H_0)$, 
essentially the same diagrammatic expansion is applicable to any classical mechanics problem with arbitrary number of particles and fields. 

Having explained the scope of the perturbation in broad terms, in the next two subsections, 
we take a closer look at the Hamiltonian deformation and the symplectic deformation, 
and show how the Magnus expansion can be applied in each case.

\subsection{Hamiltonian deformation} \label{sec:cip-hamil}

The EOM reads 
\begin{align}
    \dot{\zeta}^m  = \{ \zeta^m , H \} = \omega^{mn} \partial_n H \,.
\end{align}
The Poisson bi-vector components $\omega^{mn}$ ($\omega^{mn} \omega_{nk} = \delta^{m}_k$) are in general functions of $\zeta^m$, but we focus on the case of quadratic $\omega^\circ$, which forces $\omega^{mn}$ to be constant.  

\begin{figure}[htbp]
    \centering
    \includegraphics[width=0.4\linewidth]{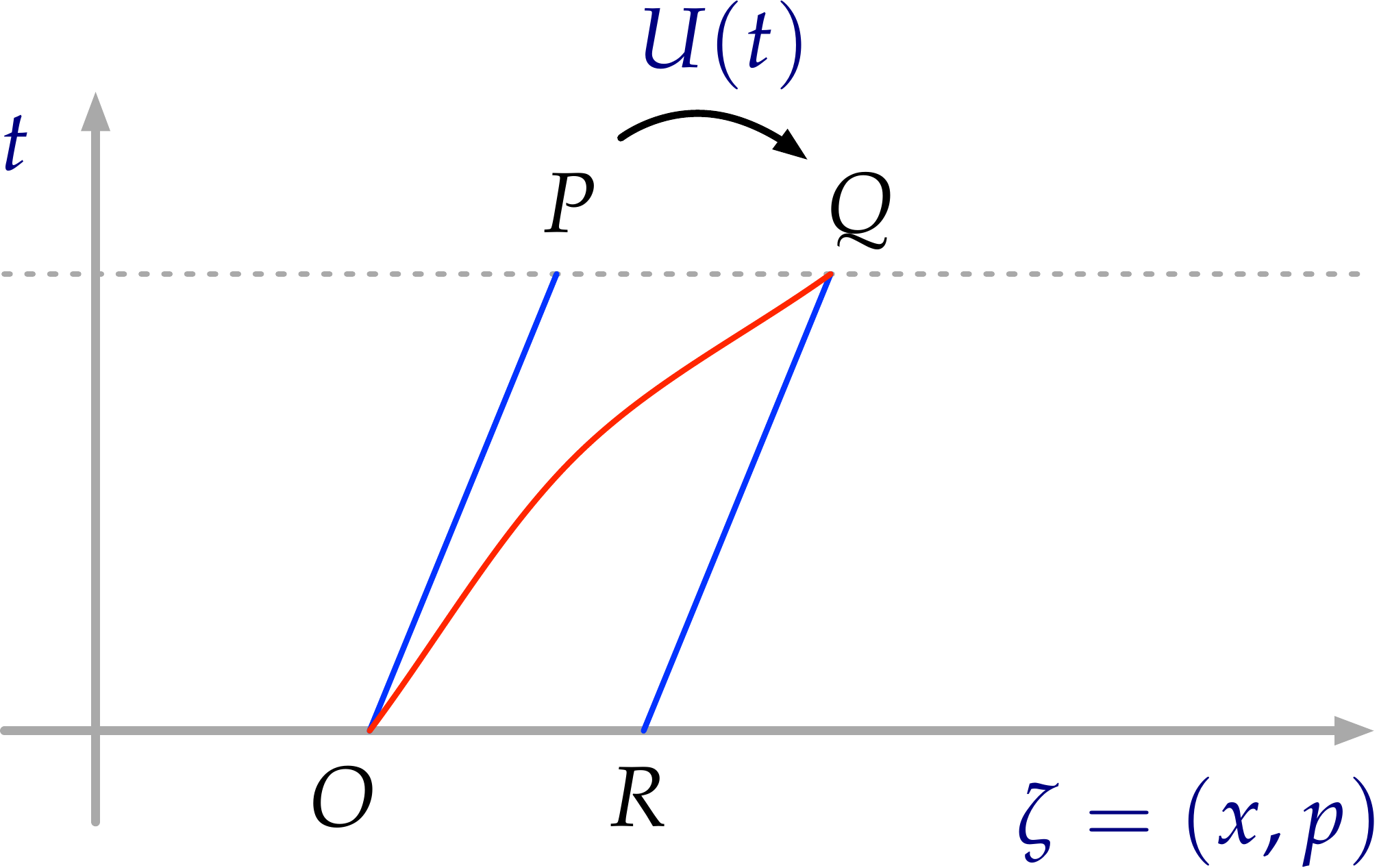}
    \caption{The classical interaction picture.}
    \label{fig:H_int}
\end{figure}

The Hamiltonian splits into two parts as in \eqref{solvable-vs-deformed}. 
We assume that both $H_0$ and $H_1$ are time-independent. 
The prime goal of the CIP is to keep track of a map between the unperturbed solution, 
\begin{align}
    \dot{\bar{\zeta}}^m(t) = \{ \bar{\zeta}^m(t), H_0(\bar{\zeta}(t)) \} \,,
\end{align}
and the perturbed solution 
\begin{align}
    \dot{\zeta}^m(t) = \{\zeta^m(t), H_0(\zeta(t)) + H_1(\zeta(t)) \} \,.
\end{align}
In Figure~\ref{fig:H_int}, the initial condition is denoted by $O$ and the unperturbed/perturbed solutions are denoted by $P$/$Q$, respectively. 
For a Hamiltonian deformation, the map between the two solutions 
stays canonical and defines the classical eikonal $\chi(t)$ as 
\begin{align}
      \zeta(t) = U(t) \bar{  \zeta}(t) = e^{\{\chi(t),\circ\}} \bar{  \zeta}(t) \,.
    \label{z-zbar-exact}
\end{align}
One of our main goals is to show that $\chi(t)$ admits Magnus expansion.

\paragraph{Review of Magnus expansion} 

We give a minimal review of the Magnus expansion following \cite{ebrahimifard2023magnusexpansion}. 
Consider a differential equation, 
\begin{align}
    \dot{Y}(t) = q A(t) Y(t) \,,
    \quad 
    Y(t_0) = Y_0
    \quad 
    \Longrightarrow
    \quad 
    Y(t)=\exp\left[ \Omega(t) \right] Y_0 \,, 
    \label{Magnus-starting}
\end{align}
where $A(t)$ and $\Omega(t)$ are elements of a Lie algebra and $Y(t)$ is an element of the representation space of the Lie algebra. $q$ is a formal coupling constant. 
Magnus~\cite{Magnus:1954zz} showed that $\Omega(t)$ is the unique solution of the differential equation, 
\begin{align}
    \dot{\Omega}(t) = q A(t) + \sum_{k>1} \frac{B_k}{k!} \operatorname{ad}^k_{\Omega}(qA)(t) = \frac{\operatorname{ad}_{\Omega}}{e^{\operatorname{ad}_{\Omega}}-1}(qA)(t) \,, 
    \quad 
    \Omega(t_0) = 0 \,.
\end{align}
Here, $\operatorname{ad}^k$ is the $k$-th iteration of $\operatorname{ad}_X(Y) := [X,Y]$ and $B_n = B_n^-$ are the Bernoulli numbers. 
The Magnus expansion in powers of $q$ is given by 
\begin{align}
    \Omega(t)=\sum_{n \geq 1} q^n \Omega_n(t) \,,
    \quad 
    \Omega_1 (t)=\int_{t_0}^t A(s) d s \,.
\end{align}
For $n\ge 2$, $\Omega_{n}$ can be computed recursively by integrating 
\begin{align}
\begin{split}
\dot{\Omega}_n(t)& = \sum_{k > 0} \frac{B_k}{k!} \sum_{\{r\}_k} [\Omega_{r_k}(t), \cdots [\Omega_{r_2}(t), [\Omega_{r_1}(t), A(t) ]] \cdots ]\,,
\end{split}
\label{Magnus-recursion}
\end{align}
where $\{r\}_k$ is the set of all $k$-tuples of integers satisfying 
$r_1 + r_2 + \cdots r_k = n-1$ and $r_i > 0$.

\paragraph{Magnus expansion for eikonal} 
Let us rewrite \eqref{z-zbar-exact} as 
\begin{align}
    \zeta(t) = U(t) \bar{\zeta}(t) = e^{-\Omega(t)} \bar{\zeta}(t) \,.
    \label{Magnus-eikonal-z(t)}
\end{align}
Taking the time derivative of both sides, 
we get 
\begin{align}
    \{ \zeta(t), H_0(\zeta(t)) + H_1(\zeta(t))\} = \frac{dU}{dt} \bar{\zeta}(t) + U\{ \bar{\zeta}(t) , H_0(\bar{\zeta}(t))\} \,.
    \label{z-zbar-full}
\end{align}
At this point, we exploit an identity, 
\begin{align}
    e^{-\Omega} \{ A, B\} = \{ e^{-\Omega} A, e^{-\Omega} B\} \,, 
    \label{eYrelation}
\end{align}
to rewrite the LHS of \eqref{z-zbar-full} as 
\begin{align}
    U \{ \bar{\zeta}(t), H_0(\bar{\zeta}(t)) + H_1(\bar{\zeta}(t))\} \,.
\end{align}
It implies an equation for $U(t)$ in terms 
of the ``interaction picture Hamiltonian" $H_I(t)$, 
\begin{align}
    \frac{d}{dt}U(t) = - U(t) X_I(t) \,,
    \quad 
    X_I(t) f = \{H_I(t), f\} \equiv \{H_1(\bar{\zeta}(t)), f\}\,.
    \label{eikonal-DE} 
\end{align}
Rephrasing it slightly, we obtain 
\begin{align}
    \frac{d}{dt} U(t)^{-1} = + X_I(t) U(t)^{-1} \,.
    \label{CIP-final}
\end{align}
Thus, we have shown that $U(t)^{-1} = e^{\Omega(t)} = e^{-\{\chi(t),\circ\}}$ admits the Magnus expansion 
with $X_I(t)$ being the ``seed" playing the role of $(qA)$ in 
\eqref{Magnus-starting}.

\paragraph{Leading order} 

To get a flavor of the Magnus expansion, we examine $\chi_{(1)}$ 
in some detail. The map between $\zeta(t)$ and $\bar{\zeta}(t)$ is approximately 
\begin{align}
    \zeta(t) = \bar{\zeta}(t) + \{\chi(t), \bar{\zeta}(t)\} + \mathcal{O}(\chi^2)\,.
\end{align}
Taking the time derivative of both sides and applying the EOM, we get 
\begin{align}
    \{\zeta, H_0(\zeta) + H_1(\zeta) \} = \{ \bar{\zeta}, H_0(\bar{\zeta})\} + \{\chi,  \{ \bar{\zeta}, H_0(\bar{\zeta})\} \} +  \{\dot{\chi}, \bar{\zeta} \} \,.
    \label{z-vs-zbar}
\end{align}
At the linearized level, the LHS can be reorganized as 
\begin{align}
    \{\zeta, H_0(\zeta) + H_1(\zeta) \} &\approx \{ \bar{\zeta} + \{ \chi,\bar{\zeta}\}, H_0(\bar{\zeta})  + \{ \chi, H_0(\bar{\zeta})\} + H_1(\bar{\zeta})  \} \,.
\end{align}
Comparing the LHS and RHS of \eqref{z-vs-zbar} 
and applying Jacobi identity to remove some terms, we obtain the linearized equation for $\chi$:
\begin{align}
    \frac{d}{dt} \chi_{(1)}(t) =  - H_I(t) 
    \quad \Longrightarrow \quad 
    - \chi_{(1)}(t) = \int_0^t H_I(s)\,ds\,. 
\end{align}

\paragraph{Impulse and Green's function} 

One use of the eikonal is to compute the impulse. 
At the leading order, the relation is 
\begin{align}
    \Delta_{(1)} f(t) = \{ \chi(t),\bar{f}(t)\} \,.
    \label{1st-impulse-general}
\end{align}
We can follow the EOM to double-check this relation. 
Setting $\zeta(t) = \bar{\zeta}(t) + \delta \zeta(t)$ and focusing on the leading fluctuation, we get 
\begin{align}
   D^m{}_n \delta \zeta^n \equiv  (\delta^m{}_n \partial_t  - \omega^{mp}\partial_n \partial_p H_0 )\delta \zeta^n = \omega^{mq} \partial_q H_1 \,, 
   \label{delta-z-EOM} 
\end{align}
where $H_0$ and $H_1$ are evaluated at $\bar{\zeta}(t)$. 
The Green's function can be derived from Poisson brackets via 
\begin{align}
\begin{split}
     &G^{mn}(t_1,t_2) = \theta(t_1-t_2) \{ \bar{\zeta}^m(t_1) , \bar{\zeta}^n(t_2) \} \,,
   \\
   &\mbox{so that} \quad 
   \{ \bar{\zeta}^m(t_1) , \bar{\zeta}^n(t_2) \}  =  G^{mn} (t_1,t_2) - G^{nm}(t_2,t_1)  \,.
\end{split}
\label{Poisson-Green}
\end{align}
The Green's function satisfies 
\begin{align}
    D^m_p G^{pn}(t_1,t_2) =  \omega^{mn} \delta(t_1-t_2) \,. 
\end{align}
Using this Green's function, we can solve the EOM \eqref{delta-z-EOM} and compute the impulse, 
\begin{align}
\begin{split}
     \Delta_{(1)} \zeta^m(t) &=  \int_0^t G^{mn}(t,s) \partial_n H_1(\bar{\zeta}(s))\, ds 
     \\
     &= \int_0^t \{ \bar{\zeta}^m(t), \bar{\zeta}^n(s)\} \partial_n H_1(\bar{\zeta}(s))\, ds
     \\
      &= \int_0^t \{ \bar{\zeta}^m(t), H_1(\bar{\zeta}(s)) \}   ds = \{ \chi_{(1)}(t) , \bar{\zeta}^m(t) \} \,,
\end{split}
\end{align}
in agreement with \eqref{1st-impulse-general}. 

\paragraph{Higher order} 

The relation between the Poisson brackets among $\bar{\zeta}^m(t)$ 
and the Green's function continues to play a crucial role 
at higher orders. 
At the 2nd order, the Magnus expansion gives
\begin{align}
\begin{split}
     \chi_{(2)} &=  \frac{1}{2} \int_0^t \theta_{12} \{ H_I(s_1), H_I(s_2) \} ds_1 ds_2 
     \\
     &= \frac{1}{2} \int_0^t \partial_m H_I(s_1) G^{mn}(s_1,s_2) \partial_n H_I(s_2) ds_1 ds_2 \,. 
\end{split}
\end{align}
The Magnus expansion itself is insensitive to the choice of $H_0$ and $H_1$. 
But, for $H_0$ that is not quadratic, some complications arise. While computing $\chi_{(3)}$, we encounter terms like
\begin{align}
  \{  H_I(s_3) , G^{ij}(s_1,s_2)\} \,.
\end{align}
When $H_0$ is quadratic, $G^{mn}$ is independent of $\zeta^m$, so this bracket vanishes. But, if $H_0$ contains cubic or higher order terms, it will produce
extra vertices in the diagrammatic expansion of the eikonal, 
in accordance with the expectation from the WQFT point of view. 
We will not discuss the issue with non-quadratic $H_0$ any further.

\paragraph{Lagrangian vs. Hamiltonian} 

We have been working in the first-order formulation that is  natural in Hamiltonian mechanics. 
But, it is often useful to ``integrate out" $p$
and work with a second-order EOM for $x$. For example, 
when $H_0 = p^2/2m + V_0(x)$ and $H_1 = V_1(x)$, 
the EOM for the leading perturbation reads
\begin{align}
\label{leading-perturbative-EOM-copy}
    \mathcal{D}_{ij} \delta x^j = \left[ m \delta_{ij} \frac{d^2}{dt^2} + \partial_i \partial_j V_0(\bar{x}(t)) \right] \delta x^j(t) = - \partial_i V_1(\bar{x}(t)) \,.
\end{align}
The Green's function for this 2nd-order differential operator is closely related to $G^{mn}(t_1,t_2)$ discussed above. 
The Green's function restricted to the $x$-space, 
\begin{align}
    \mathcal{G}^{ij}(t_1,t_2) = \theta(t_1-t_2) \{ \bar{x}^i(t_1), \bar{x}^j(t_2) \} \,,
    \label{Green-xx}
\end{align}
is easily shown to satisfy  
\begin{align}
    \mathcal{D}_{ij} \mathcal{G}^{jk}(t_1,t_2) = - \delta_i{}^k \delta(t_1-t_2) \,. 
\end{align}

\subsection{Symplectic deformation} \label{sec:cip-symp}

The symplectic form splits into two parts as in \eqref{solvable-vs-deformed}.
Generically, the deformed Poisson bracket gives rise to an infinite series, 
\begin{align}
    \{f, g\}=\{f, g\}_{\circ} - \{f, \zeta^m\}_{\circ} \omega_{mn}^{\prime} \{\zeta^n, g \}_{\circ}
    + \{f, \zeta^m\}_{\circ} (\omega' \omega_\circ \omega')_{mn} \{\zeta^n, g \}_{\circ}
    +\mathcal{O}(\omega')^3 \,.
    \label{symp-inversion}
\end{align}
Here, $(\omega_\circ)^{mn}$ denotes the inverse matrix of $\omega^\circ_{mn}$ and  
$(\omega' \omega_\circ \omega')_{mn} = \omega'_{mp} (\omega_\circ)^{pq} \omega'_{qn}$.

Again we should connect the unperturbed solution, 
\begin{align}
    \dot{\bar{\zeta}}^m(t) = \{ \bar{\zeta}^m(t), H(\bar{\zeta}(t)) \}_\circ \,,
\end{align}
to the perturbed solution, 
\begin{align}
    \dot{\zeta}^m(t) = \{\zeta^m(t), H(\zeta(t)) \}_{\zeta(t)} \,.
\end{align}
The notation $\{ \; , \;\}_{\zeta(t)}$ emphasizes the fact that the deformed Poisson bracket acquires time dependence through $\zeta(t)$, whereas the undeformed bracket is time-independent (assuming a quadratic $\omega^\circ$). 
Assuming a map between the two as before, 
\begin{align}
      \zeta(t) = U(t) \bar{\zeta}(t) \,.
\end{align}
and comparing the EOM for the two solutions, we obtain a differential equation for $U(t)$:
\begin{align}
    U^{-1} \frac{dU}{dt} f= \{ f, H(\bar{\zeta}(t)) \}_{\bar{\zeta}(t)} -  \{ f, H(\bar{\zeta}(t)) \}_\circ \,.
    \label{symp-CIP}
\end{align}
The first term on the RHS uses the deformed bracket evaluated along the unperturbed trajectory. 

\paragraph{1st order eikonal} 

In a scattering problem where the eikonal is well defined,
the 1st order eikonal is given by 
\begin{align}
    \chi_{(1)} = \int \theta'_m(\bar{\zeta}(s)) \dot{\bar{\zeta}}^m(s) ds \,, 
    \quad 
    \omega' = d\theta' \,.
    \label{chi1-symp}
\end{align}
Let us check whether it reproduces the 1st order impulse. 
The EOM for the 1st order perturbation is 
\begin{align}
    (\delta^m{}_n \partial_t  - \omega^{mp}_\circ \partial_n \partial_p H)\delta z^n = - \,(\omega_\circ \omega' \omega_\circ) ^{mq} \partial_q H = - \,(\omega_\circ \omega')^m{}_p \dot{\bar{\zeta}}^p\,.
   \label{delta-z-EOM-symp} 
\end{align}
Using the same Green's function as before, we can express the impulse as 
\begin{align}
    \Delta_{(1)} \zeta^m(t) = -\int_0^t \{\bar{\zeta}^m(t), \bar{\zeta}^n(s) \}_\circ \omega'_{np} \dot{\bar{\zeta}}^p(s) ds \,.
\end{align}
Using $\omega'_{np} = \partial_n \theta'_p - \partial_p \theta'_n$, we get 
\begin{align}
\begin{split}
       &\int_0^t \{\bar{\zeta}^m(t), \bar{\zeta}^n(s) \}_\circ (\partial_n \theta'_p - \partial_p \theta'_n) \dot{\bar{\zeta}}^p ds
       \\
       &=\int_0^t \left[ \{\bar{\zeta}^m(t), \theta'_p(\bar{\zeta}(s)) \}_\circ \dot{\bar{\zeta}}^p(s) - \{\bar{\zeta}^m(t), \bar{\zeta}^n(s) \}_\circ \frac{d}{ds} \theta'_n(\bar{\zeta}(s)) \right] ds
       \\
       &\approx \int_0^t \left[ \{\bar{\zeta}^m(t), \theta'_p(\bar{\zeta}(s)) \}_\circ \dot{\bar{\zeta}}^p(s) + \{\bar{\zeta}^m(t), \dot{\bar{\zeta}}^n(s) \}_\circ  \theta'_n(\bar{\zeta}(s)) \right] ds
       \\
       &= \int_0^t \{\bar{\zeta}^m(t), \theta'_p(\bar{\zeta}(s)) \dot{\bar{\zeta}}^p(s) \} ds\,, 
\end{split}
\end{align}
which proves \eqref{chi1-symp} when the total derivative vanishes. 

\paragraph{2nd order eikonal} 

The WQFT Feynman rules readily suggests that 
\begin{align}
    \chi_{(2)} = \frac{1}{2} \int G^{mn}(s_1,s_2) \omega'_{mp}(s_1) \dot{\bar{\zeta}}^p(s_1) \omega'_{nq}(s_2) \dot{\bar{\zeta}}^q(s_2) ds_1 ds_2 \,.
    \label{symp-chi2}
\end{align}
But, reproducing it on the Magnus side is slightly nontrivial. 

Combining the infinite series for the deformed Poisson bracket \eqref{symp-inversion} 
and the defining equation for the CIP with symplectic deformation \eqref{symp-CIP}, 
we find that an infinite series of vector fields contribute to the Magnus expansion, 
\begin{align}
\begin{split}
       X_1(s) &= (\omega_\circ \omega')^m{}_n \dot{\bar{\zeta}}^n(s) \partial_m \,, 
       \\
       X_2(s) &= - (\omega_\circ \omega'\omega_\circ \omega')^m{}_n \dot{\bar{\zeta}}^n(s) \partial_m \,, 
        \\
       X_3(s) &=  (\omega_\circ \omega'\omega_\circ \omega'\omega_\circ \omega')^m{}_n \dot{\bar{\zeta}}^n(s) \partial_m \,, 
       \quad \mbox{etc.}
\end{split}
\end{align}
The vector field relevant for $\chi_{(2)}$ is the sum of two terms, 
\begin{align}
- \{ \chi_{(2)}, f \} =  \left[  \frac{1}{2} \int \theta_{12} [ X_1(s_1) , X_1(s_2)] ds_1 ds_2 + \int X_2(s) ds  \right] f\,.
\label{symp-magnus2}
\end{align}
After some cancellations, we can verify that \eqref{symp-chi2} follows from \eqref{symp-magnus2} 
provided that some total derivatives vanish. 

Comparing the Magnus expansion with the WQFT Feynman rules is straightforward in principle but cumbersome in practice.  
In section~\ref{sec:EM-background}, in a relatively simple setting, 
we will verify to all orders in perturbation theory that the Magnus expansion gives the same diagrammatic expansion as the WQFT Feynman rules.

\subsection{Particles} 

The discussions in previous subsections have been general but rather abstract. 
In this and the next subsections, we give some concrete examples 
where the solutions to the solvable (free) theory does not necessarily represent straight line trajectories. 

\paragraph{Anharmonic oscillator} 
As an example where $H_0$ includes some interaction, consider a simple harmonic oscillator perturbed by an anharmonic term: 
\begin{align}
    H_0 = \frac{p^2}{2m} + \frac{1}{2} m\omega^2 x^2\,,
    \quad 
    H_1 = \frac{\lambda}{4!} x^4 \,.
\end{align}
The interaction-picture Hamiltonian is 
\begin{align}
      H_I(x,p;t) = H_1(\bar{x}(t),\bar{p}(t))= \frac{\lambda}{4!} \left( x \cos\omega t + \frac{p}{m\omega} \sin\omega t \right)^4 \,.
\end{align}
The map from the unperturbed trajectory to the perturbed one \eqref{z-zbar-exact} reads 
\begin{align}
\begin{split}
    x(t) &= e^{\{\chi(t),\circ\}} \left(x \cos\omega t + \frac{p}{m\omega} \sin\omega t \right) \,,
    \\
    p(t) &= e^{\{\chi(t),\circ\}} \left(p \cos\omega t - x(m\omega)  \sin\omega t \right)  \,.    
\end{split}
\end{align}
The retarded Green's function for the differential operator $\partial_t^2 + \omega^2$ is $\theta(t_1 - t_2) \sin(t_1-t_2)$. It is straightforward to compute $\chi_{(n)}(t)$ at an arbitrary time $t$ to many orders.

\paragraph{Impulse in two steps} 

Given an initial condition, $\zeta(0)$, in view of Figure~\ref{fig:H_int}, we may define 
three notions of ``impulse": 
\begin{align}
    \Delta_{PO}(\zeta) = \bar{\zeta}(t) - \zeta(0) \,,
    \quad 
    \Delta_{QO}(\zeta) = \zeta(t) - \zeta(0) \,,
    \quad 
    \Delta_{QP}(\zeta) = \zeta(t) - \bar{\zeta}(t) \,.
\end{align}
In the interacting theory with $H=H_0+H_1$, our final goal is $\Delta_{QO}(\zeta)$. Since we cannot solve the full theory exactly, 
we use the CIP to compute it in two steps: $\Delta_{QO} = \Delta_{QP} + \Delta_{PO}$. 
There are two special cases where the ``solvable" part ($\Delta_{PO}$) is particularly simple.

\paragraph{Scattering} The trajectory asymptotes to straight lines in the past/future infinities. 
The scattering observables are $p_\mu$ and the transverse position $b^\mu = x^\mu - p^\mu (x\cdot p)/p^2$. 
Both $p_\mu$ and $b^\mu$ remain unchanged along the free straight-line trajectory.

\paragraph{Periodic orbit}

The unperturbed trajectory comes back to the same point in the phase space after a time period, $T$, 
so that $\Delta_{PO}(z)|_{t=T} = 0$. 
The perturbed trajectory may deviate from the periodic orbit (cf. perihelion precession). 
The anharmonic oscillator above is particularly simple in that the period of the unperturbed orbit is independent of the initial condition: $T= 2\pi/\omega$. 
The eikonal tends to simplify when $t$ is an integer multiple of the period. 
For example, the 1st order eikonal at $t=T = 2\pi /\omega$ is 
\begin{align}
    \chi_{(1)}(T) = -\frac{T}{32} \left[x^2+ \left(\frac{p}{m\omega}\right)^2\right]^2 \,.
\end{align}

A similar analysis should be possible, but would be more involved, for more general periodic orbits. 
An important application is the post-Newtonian expansion \eqref{PN-Hamiltonian}.
We leave the study of the PN expansion based on the classical eikonal for a future work.

\subsection{Fields} \label{sec:CIP-fields}

Classical field theory is classical mechanics with infinitely many degrees of freedom. In particular, a free field theory describes a set of infinitely many coupled harmonic oscillators. 
To be specific, consider a Klein-Gordon field coupled to a source:
\begin{align}
    \mathcal{L} = -\frac{1}{2} \partial^\mu \phi \partial_\mu \phi - \frac{1}{2} m^2 \phi^2 + J\phi \,.
\end{align}
The EOM is 
\begin{align}
    (-\partial^\mu \partial_\mu + m^2) \phi = J \,.
\end{align}
The inhomogeneous solution is 
\begin{align}
    \phi(y) = \int d^4z \,G(y-z) J(z)  \,,
    \label{KG-solution}
\end{align}
where $G$ is the retarded propagator,
\begin{align}
\begin{split}
    \quad G(x) &=  \int \bar{d}^4 k \frac{e^{ik\cdot x}} {k^2 - \mathrm{sgn}(k^0) i \epsilon+m^2} 
    \\
    &=  \theta(t) \int \bar{d}^3 k \frac{(+i)}{2\omega_k} e^{i\vec{k}\cdot \vec{x} } \left(  e^{-i\omega_k t} - e^{+i\omega_k t} \right) 
    \\
    &= (+i) \theta(t)  \int \widetilde{dk} ( e^{ik\cdot x} - e^{-ik\cdot x})\,.    
\end{split}
\label{G_R-massive}
\end{align}

We can rephrase the solution \eqref{KG-solution} in terms of a classical eikonal.
Before turning on the source $J$, the time evolution of the field is well known. 
We can quote the results from QFT textbooks, except that we work with Poisson brackets rather than quantum commutators. 
\begin{align}
\begin{split}
    \phi(t,\vec{x}) &= \int \widetilde{dk} \left[ a_k e^{ik\cdot x} + a_k^* e^{-ik\cdot x} \right]\,,   
    \\
     \pi(t,\vec{x}) &= (-i) \int \widetilde{dk} \,\omega_k \left[ a_k e^{ik\cdot x} - a_k^* e^{-ik\cdot x} \right]\,.
\end{split}
\label{KG-mode-expansion}
\end{align}
The modes $a_k$, $a_k^*$ satisfy 
\begin{align}
    \{ a_p, a^*_q \} = (-i) (2\omega_p)\bar{\delta}^3(\vec{p} -\vec{q}) \,.
\end{align}
The free Hamiltonian,  
\begin{align}
    H_0 = \int \widetilde{dk} \, \omega_k (a_k^* a_k) \,,
\end{align}
generates the time-evolution in the sense that 
\begin{align}
    \begin{pmatrix}
        \phi(t,\vec{x}) \\ \pi(t, \vec{x})
    \end{pmatrix} = 
    e^{-t \{H_0, \circ\} } 
    \begin{pmatrix}
        \phi(0,\vec{x}) \\ \pi(0, \vec{x}) 
    \end{pmatrix} \,.
\end{align}
The Poisson bracket of the free field is related to the propagators as 
\begin{align}
    \{ \phi(x), \phi(y) \} 
    = -G(x-y) + G(y-x) \equiv D(x-y)\,.
    \label{R-A-KG}
\end{align}
This relation is a field theory counterpart of \eqref{Poisson-Green}. 
The combination of \eqref{Poisson-Green} and \eqref{R-A-KG} was called a ``causality" cut in \cite{Kim:2024svw}. 
In a scattering problem, when one of the coordinates, say $y$ in \eqref{R-A-KG}, is taken at the ``future infinity" 
while other coordinates, say $x$, are integrated over a region, we will assume that $y^0 > T > x^0$ and take $T \rightarrow + \infty$ only at the final stage. 

The source term in the Hamiltonian is 
\begin{align}
H_1(s) = - \int d^3\vec{x} \, J(s,\vec{x}) \phi(s,\vec{x}) \,.
\end{align}
It translates to the interaction Hamiltonian, 
\begin{align}
    H_I(s) = - \int d^3\vec{x} \, J(s,\vec{x}) \phi(s,\vec{x}) \,,
\end{align}
where $\phi(s,\vec{x})$ inside the integral is understood as the mode expansion \eqref{KG-mode-expansion}. 
The first order eikonal $\chi_{(1)}$ recovers Lorentz covariance, 
\begin{align}
    \chi_{(1)} =  - \int dt\,H_I(t)  = \int d^4x \, J(x) \phi(x) \,.
\end{align}
The change of the field configuration produced by the source is 
\begin{align}
\begin{split}
       \Delta_{(1)} \phi(y) &= \{ \chi_{(1)}, \phi(y) \} 
       = \left\{ \int d^4y\,J(x) \phi(x) ,  \phi(y) \right\} 
       = \int d^4 y \,G(y-x) J(x) \,,
\end{split}
\end{align}
where we used \eqref{R-A-KG} and assumed that $y^0 > x^0$. 
We observe that the first order ``impulse" already reproduces the exact solution \eqref{KG-solution}. 
That is because $H_I$ is linear in $\phi$. It is easy to see that $\{H_I(t), H_I(s)\}$ is independent of $\phi$, 
so that $\{ \chi_{(n)} , \phi \} = 0$ for all $n\ge 2$. 

\paragraph{Scalar field}
Let us proceed to a more general setting and consider 
\begin{align}
    \mathcal{L} = -\frac{1}{2} \partial^\mu \phi \partial_\mu \phi - \frac{1}{2} m^2 \phi^2 - V(\phi, J) \,,
\end{align}
where $J$ symbolically represents the set of all possible external variables. The EOM is 
\begin{align}
    (-\partial^\mu \partial_\mu +m^2) \phi = -\frac{\partial V}{\partial \phi} \,.
\end{align}
We solve the EOM perturbatively around a background by setting $\phi = \bar{\phi} + \delta\phi$. 
The background field $\bar{\phi}$ is a homogeneous solution to the free KG equation (linear combination of plane waves). 
So, the EOM becomes
\begin{align}
      (-\partial^\mu \partial_\mu +m^2) \delta \phi = - \partial_\phi V(\bar\phi+\delta\phi, J) \,.
\end{align}
The structure of the EOM is essentially the same as that of 
the anharmonic oscillator. 
So, we can treat the particles and fields on an equal footing 
and apply the Magnus expansion to the whole system in a uniform way. 
For a self-interacting scalar field, the formula for perturbative eikonal 
clearly shows the parallel between particles and fields. For example, 
\begin{align}
\begin{split}
     - \chi_{(1)} &=  \int V(\bar{\phi}(x_1)) \,d^4 x_1 \,,
     \\
     - \chi_{(2)} &= - \frac{1}{2} \int \partial V(\bar{\phi}(x_1))  G(x_1-x_2) V(\bar{\phi}(x_2))\, d^4x_1 d^4 x_2 \,. 
     \label{chi12-KG}
\end{split}
\end{align}
The ``impulses" are computed in the by now familiar way, 
\begin{align}
\begin{split}
       \Delta_{(1)} \phi(y) &= \{ \chi_{(1)} , \phi(y) \}  \,, 
       \\
       \Delta_{(2)} \phi(y) &=  \{ \chi_{(2)} , \phi(y) \}   +  \frac{1}{2} \{ \chi_{(1)} ,  \{ \chi_{(1)} , \phi(y) \}   \}  \,, 
\end{split}
\end{align}

Even for a scalar field, the Hamiltonian formulation introduces $(1+3)$ slicing of the Minkowski spacetime 
and the Lorentz symmetry is slightly obscured. The Lorentz invariance is restored in the end,  
as we can see in \eqref{chi12-KG}. 
Later in this paper, we will deal with photon and graviton fields. 
The U$(1)$ gauge symmetry of the photon field and the general covariance of the graviton field 
cause some complications in the Hamiltonian formulation, but 
for most practical computations will only need generalizations of the relation \eqref{R-A-KG} between the Poisson bracket and the Green's function .

\paragraph{Photon field}

The Maxwell Lagrangian coupled to an external current is 
\begin{align}
    \mathcal{L} = - \frac{1}{4} F_{\mu\nu} F^{\mu\nu} + A_\nu J^\nu \,.
\end{align}
The Maxwell equation reads 
\begin{align}
    \partial_\mu F^{\mu\nu} + J^\nu = 0 \,.
\end{align}
In Lorenz gauge, the solution to the initial value problem is 
\begin{align}
    A_\mu(y) = \int d^4 y \, G(y-z) J_\mu(z) \,,
\end{align}
where $G$ is the massless limit of the retarded propagator \eqref{G_R-massive}. 
Whenever possible, it is desirable to work with the gauge-invariant field strength, 
\begin{align}
       F_{\mu\nu}(y) = \int d^4 z \left[ \partial_\mu G(y-z) J_\nu(z) - \partial_\nu G(y-z) J_\mu(z) \right]\,. 
       \label{F-from-current}
\end{align}
The covariant Poisson bracket \eqref{R-A-KG} generalizes to 
\begin{align}
    \{ A_\mu(x), A_\nu(y) \} = \eta_{\mu\nu} D(x-y) \,. 
    \label{AA-bracket}
\end{align} 
Turning one of the potentials to a field strength, we have
\begin{align}
      \{ F_{\mu\nu}(x), A_\rho(y) \} &= \partial_\mu D(x-y) \eta_{\nu\rho} 
      -  \partial_\nu D(x-y) \eta_{\mu\rho} \,.
\label{FA-bracket}
\end{align}
Similarly, the bracket between two field strengths is 
\begin{align}
\begin{split}
      \{ F_{\mu\nu}(x), F_{\rho\sigma}(y) \} &= -\partial_\mu \partial_\rho D(x-y) \eta_{\nu\sigma} 
      +\partial_\nu \partial_\rho D(x-y) \eta_{\mu\sigma}
      \\
      &\qquad +  \partial_\mu \partial_\sigma D(x-y) \eta_{\nu\rho} - \partial_\nu \partial_\sigma D(x-y) \eta_{\mu\rho}
      \,. 
\end{split}
\label{FF-bracket}
\end{align}
The same discussion applies to the graviton field, but we will not need it in this paper.

 
\section{Electromagnetic background} \label{sec:EM-background}

The EOM for a charged particle in an electromagnetic (EM) field background is  
\begin{align}
    \begin{split}
        \frac{dx^\mu}{d\tau } &= \frac{p^\mu}{m}  \,,
        \\
         \frac{dp_\mu}{d\tau } &= \frac{q}{m} F_{\mu\nu} p^\nu = q F_{\mu\nu} \dot{x}^\nu  \,.
    \end{split}
    \label{EM-EOM} 
\end{align}
It is one of the simplest example of symplectic perturbation, 
\begin{align}
H =  \frac{p^2+m^2}{2m}\,,
\qquad 
\omega = \omega^\circ + \omega^q = dx^\mu \wedge dp_\mu + \frac{q}{2} F_{\mu\nu} dx^\mu \wedge dx^\nu \,.
\end{align}
The Poisson bracket is deformed accordingly, 
\begin{align}
    \{ f , g\} =  \{ f , g\}_\circ +  \{ f , g\}_{q} = \frac{\partial f}{\partial x^\mu} \frac{\partial g}{\partial p_\mu}  - \frac{\partial f}{\partial p_\mu}  \frac{\partial g}{\partial x^\mu}  + q F_{\mu\nu}(x)  \frac{\partial f}{\partial p_\mu}  \frac{\partial g}{\partial p_\nu} \,. 
\end{align}
The Bianchi identity for $F_{\mu\nu}$ implies $d\omega = 0$, which in turn ensures that the deformed bracket satisfies Jacobi identity. 

Compared to the general symplectic deformation discussed in section~\ref{sec:cip-symp}, 
the current problem is simpler in a few ways. 
First, the Hamiltonian describes a free particle. 
(More precisely, $H=(\kappa/2)(p^2+m^2)$ comes with a Lagrange multiplier $\kappa$ enforcing the mass shell condition. 
We are working in the gauge where $\kappa=1/m$.) 
Second, the deformed bracket in general consists of an infinite sum, but in this example, it truncates exactly at the 1st order in $q$. 
Rather than importing the formulas from section~\ref{sec:cip-symp}, 
we find it instructive to develop the CIP step by step in this simple setting. 
Our starting point is 
\begin{align}
\begin{split}
     x^\mu(\tau) &= U(\tau) \bar{x}^\mu(t) = U(\tau)(x^\mu +v^\mu\tau) \,,
     \\
     p_\mu(\tau) &= U(\tau) \bar{p}_\mu(t) = U(\tau)p_\mu\,. 
\end{split}
\end{align}
Applying the EOM \eqref{EM-EOM} on the LHS and using the ansatz 
$dU(\tau)/d\tau = - U(\tau)X_I(\tau)$, we find two conditions for $X_I(\tau)$, 
\begin{align}
\begin{split}
        &-X_I(\tau)(x^\mu + v^\mu\tau) = 0 \,,
    \\
        &-X_I(\tau)p_\mu = q E_{\mu}(x+v\tau) \,,
\end{split}
\end{align}
where we introduced a shorthand notation for the ``electric field vector":
\begin{align}
  E_\mu(\tau) \equiv    F_{\mu\nu}(\tau) v^\nu \,, \quad 
  v^\nu \equiv \frac{p^\nu}{m} \,.
\end{align}
We observe that the Magnus expansion works with the vector field
\begin{align}
    - X_I(\tau) =  E_\mu(x+v\tau) \left(  \frac{\partial}{\partial p_\mu} - \frac{\tau}{m}  \frac{\partial}{\partial x_\mu}\right) \equiv E_\mu D^\mu \,.
\end{align}
For a scattering process where $F_{\mu\nu}(b+v\tau) \rightarrow 0$ as $\tau \rightarrow \pm \infty$, 
we will show how to ``extract" the classical eikonal $\chi$.
But, for a finite time interval problem or a bound orbit (e.g. cyclotron motion), 
unless the time interval equals the period of the motion, 
we cannot turn the vector field into a scalar generator $\chi$ 
in a gauge-invariant way. 
From the Newtonian point of view, 
the electric field $E_\mu$ behaves like $\nabla V$.

\subsection{Eikonal up to the 3rd order}

\paragraph{1st order} 

The 1st order impulse is simply
\begin{align}
\Delta_{(1)} p_\mu  = q \int F_{\mu\nu}(\tau) v^\nu d\tau  = q \int E_{\mu}(\tau)\, d\tau \,.
\label{1st-impulse-EM}
\end{align}
The 1st order eikonal generating this impulse can be written as
\begin{align}
    \chi_{(1)} = q \int v^\mu A_\mu(x+v\tau)\, d\tau \,. 
    \label{chi1-EM}
\end{align}
A gauge transformation, $A_\mu \rightarrow A_\mu + \partial_\mu \Lambda$, produces a total derivative in the integrand, which vanishes upon integration under the assumption that the gauge field falls off asymptotically. 
To see how $\chi_{(1)}$ acts on observables, we note that 
\begin{align}
    \{ A_\mu(x+v\tau) v^\mu, f \} = - X(\tau) f + \left(\frac{dA_\mu}{d\tau}  \right)\frac{\partial f}{\partial p_\mu} - \frac{1}{m} \frac{d}{d\tau}(\tau A_\mu) \frac{\partial f}{\partial x_\mu} \,.
    \label{total-derivative-1}
\end{align}
When $O$ is independent of $\tau$, the second and third terms are total derivatives, so we get
\begin{align}
    \{ \chi_{(1)}, f \} = \left( -q\int X(\tau) d\tau \right) f \,, 
\end{align}
reproducing \eqref{1st-impulse-EM}. 
From here on, we write $X(\tau)$ in place of $X_I(\tau)$ and $X_i \equiv X(\tau_i)$.

\paragraph{2nd order} 
$\chi_{(2)}$ is related to the vector field
\begin{align}
 - \Omega_{(2)} = \frac{q^2}{2} \int \theta_{12}  [ X(\tau_1) , X(\tau_2) ] d\tau_1 d\tau_2 \,.
\end{align}
The Lie bracket consists of two terms, 
\begin{align}
\begin{split}
      [ X(\tau_1) , X(\tau_2) ] &= E_\mu(\tau_1) \left[ D_1^\mu E_\nu(\tau_2)\right] D_2^\nu -E_\mu(\tau_2) \left[ D_2^\mu E_\nu(\tau_1)\right] D_1^\nu 
      \\
      &= \frac{1}{m} E_\mu(\tau_1) \left[ F_\nu{}^\mu(\tau_2) + \tau_{21} \partial^\mu E_\nu(\tau_2) \right] D_2^\nu - (1\leftrightarrow 2) \equiv \frac{1}{m} (W_2+W_1) \,.
\end{split}
\end{align}
Applying the Bianchi identity in the form, 
\begin{align}
    \begin{split}
        &v^\rho \left[ \partial_\rho F_{\mu \nu}(\tau) + \partial_\mu F_{\nu\rho}(\tau) + \partial_\nu F_{\rho \mu}(\tau)\right] = 0 
        \\
        &\Longleftrightarrow\quad  \frac{d}{d\tau} F_{\mu\nu}(\tau) + \partial_\mu E_\nu(\tau) - \partial_\nu E_\mu(\tau) = 0 \,, 
    \end{split}
    \label{Bianchi-E}
\end{align}
we obtain
\begin{align}
      W_2 
      = E_\mu(\tau_1) \left[ \tau_{21} \partial_\nu E^\mu(\tau_2)   + \frac{d}{d\tau_2} \left( \tau_{21} F_\nu{}^\mu(\tau_2) \right) \right] D_2^\nu  \,.  
\label{Lie-12-simpler}
\end{align}
Up to a total derivative in the $\tau_2$ integral, we obtain 
\begin{align}
    W_2 \approx -\tau_{12} E_\mu(\tau_1)\left[  \partial_\nu E^\mu(\tau_2) D_2^\nu - \frac{1}{m} F^{\mu\nu} (\tau_2) \frac{\partial}{\partial x^\nu}\right]  \,.
\end{align}
Similarly, we find 
\begin{align}
    W_1 \approx -\tau_{12} E_\mu(\tau_2)\left[  \partial_\nu E^\mu(\tau_1) D_1^\nu - \frac{1}{m} F^{\mu\nu} (\tau_1) \frac{\partial}{\partial x^\nu}\right]  \,.
\end{align}

We are ready to write down the 2nd order eikonal $\chi_{(2)}$ that generates $\Omega_{(2)}$: 
\begin{align}
    \{ \chi_{(2)}, O \} = \Omega_{(2)} O \,.
    \label{chi-Omeaga-2-EM}
\end{align}
The precise form of $\chi_{(2)}$ is given by
\begin{align}
    \chi_{(2)} = \frac{q^2}{2m} \int R_{12}\left[  E_{\mu}(\tau_1)  E^{\mu}(\tau_2 ) \right]  d\tau_1 d\tau_2 \,.
    \label{chi2-EM}
\end{align}
To verify \eqref{chi-Omeaga-2-EM}, it suffices to use the identity,
\begin{align}
    \{E^\mu(\tau_i) ,O\} = \partial_\nu E^\mu(\tau_i) D_i^\nu O -\frac{1}{m} F^{\mu\nu}(\tau_i) \frac{\partial}{\partial x^\nu} O \,, 
    \label{E-bracket-O}
\end{align}
which follows from $D_i^\mu(x^\nu+v^\nu \tau_i)=0$ and the identity, 
\begin{align}
\begin{split}
       \{g, h\} &= \frac{\partial g}{\partial x^\mu} \frac{\partial h}{\partial p_\mu} - \frac{\partial g}{\partial p_\mu} \frac{\partial h}{\partial x^\mu} 
    = \frac{\partial g}{\partial x^\mu} \left(\frac{\partial h}{\partial p_\mu} - \frac{\tau}{m} \frac{\partial h}{\partial x_\mu} \right)- \left(\frac{\partial g}{\partial p_\mu} - \frac{\tau}{m} \frac{\partial g}{\partial x_\mu} \right)\frac{\partial h}{\partial x^\mu} 
    \\
    &= (\partial_\mu g )(D^\mu h) - (D^\mu g) (\partial_\mu h) \,.
\end{split}
\label{Poisson-D-shift}
\end{align}

\paragraph{3rd order} 

Again, we should first compute 
\begin{align}
   - \Omega_{(3)} = \frac{q^3}{6} \int \theta_{12} \theta_{23} 
   \left( [X_1, [X_2, X_3]] + [X_3, [X_2, X_1]] \right) d\tau_1 d\tau_2 d\tau_3 \,, 
   \label{Omega3-X}
\end{align}
and find a scalar function that generates $\Omega_{(3)}$. 
The procedure is similar to the one for $\chi_{(2)}$.

\begin{figure}[htbp]
    \centering
    \includegraphics[width=0.55\linewidth]{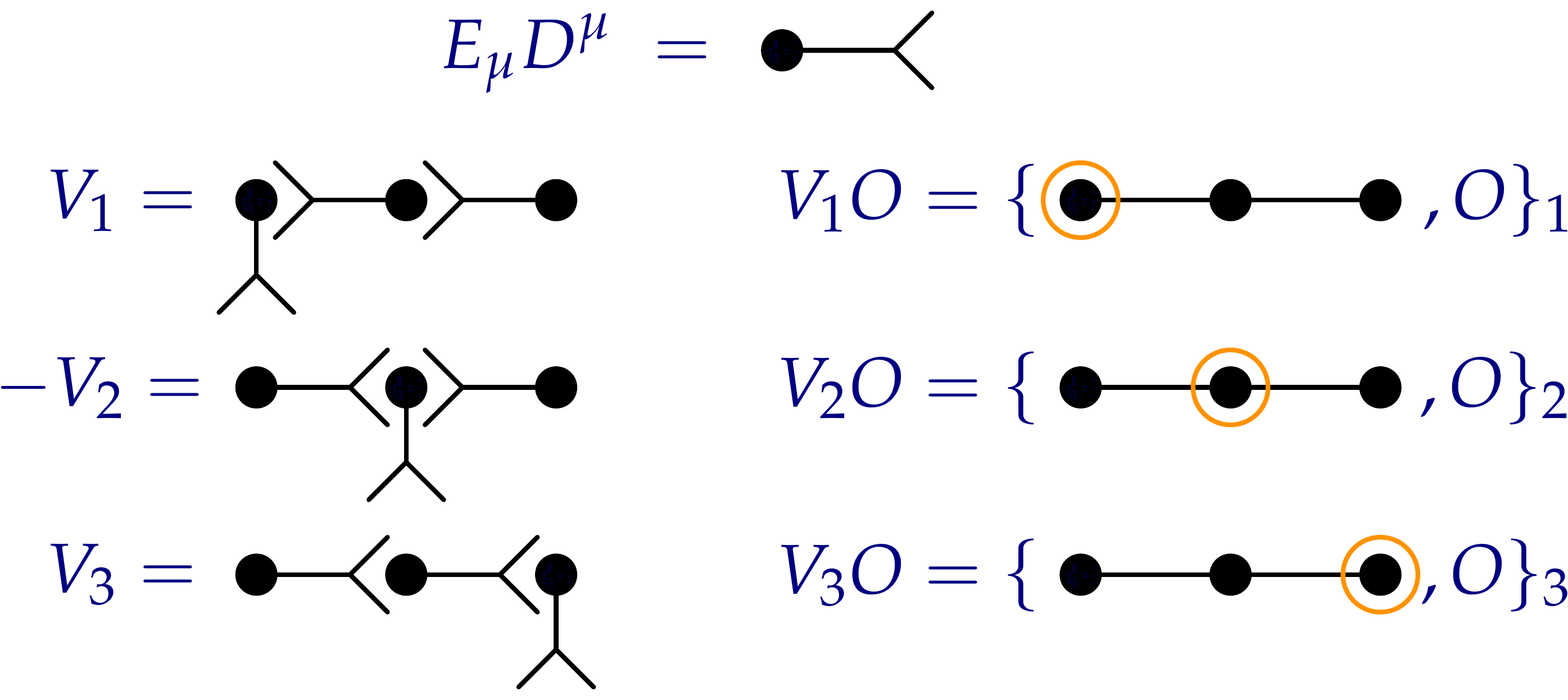}
    \caption{A graphical notation for composition of vector fields.}
    \label{fig:extraction-chi3}
\end{figure}

The Lie brackets in \eqref{Omega3-X} produce many terms, but we don't have to compute them all at the same time, since trees of different topology do not mix. 
For example, the ``linear" tree at the 3rd order corresponds to the vector field, 
\begin{align}
\begin{split}
&V = V_1+ V_2 + V_3 \,,
\\
   V_1 &= \int \theta_{12} \theta_{23} E_\nu(\tau_3)[D_3^\nu E_\mu(\tau_2)] [D_2^\mu E_\rho(\tau_1)] D_1^\rho  \,,
     \\
     V_2 &= -   \int \theta_{12} \theta_{23} 
    E_\mu(\tau_1) E_\nu(\tau_3) [D_1^\mu D_3^\nu E_\rho(\tau_2)] D_2^\rho  \,,
   \\
   V_3 &=\int \theta_{12} \theta_{23} E_\mu(\tau_1)[D_1^\mu E_\nu(\tau_2)] [D_2^\nu E_\rho(\tau_3)] D_3^\rho \,.
\end{split}
\label{V3-linear}
\end{align}
As sketched in Figure~\ref{fig:extraction-chi3}, 
the relation $VO = \{f, O\}$ follows from the refined relations, 
\begin{align}
    V_i O = \{ f, O \}_i  
    \qquad (i =1,2,3)
    \label{VO-fO}
\end{align}
where the Poisson bracket with a label $\{ \;\; , \;\; \}_i$ means that the derivatives act on the $i$-th vertex of $f$ only. 
Since $V_i$ ($i=1,2,3$) are all similar, 
it suffices to consider, say, $V_3$.
\begin{align}
\begin{split}
   V_3 =\int \theta_{12} \theta_{23} E_\mu(\tau_1)[D_1^\mu E_\nu(\tau_2)] [D_2^\nu E_\rho(\tau_3)] D_3^\rho \,,
\end{split}
\end{align}
In what follows, $E_\mu(\tau_1)$ acts like a spectator. Using the computations leading to \eqref{E-bracket-O}, we can process $[D_2^\nu E_\rho(\tau_3)] D_3^\rho$ to find
\begin{align}
    V_3 O = - \frac{1}{m} \int \theta_{12} R_{23} E_\mu(\tau_1)[D_1^\mu E_\nu(\tau_2)] \{ E^\nu(\tau_3) , O \} \,.
\end{align} 
Compared to the computation for $\chi_{(2)}$, the nontrivial part is $[D_1^\mu E_\nu(\tau_2)]$. We write it as \begin{align}
    D_1^\mu E^\nu(\tau_2) = \frac{1}{m}\left[ -\tau_{12} \partial^{(\mu} E^{\nu)}(\tau_2) + \frac{1}{2} \tau_{12} \frac{d}{d\tau_2} F^{\mu\nu}(\tau_2) - F^{\mu\nu}(\tau_2) \right]
\end{align}
After an IBP in $\tau_2$, we confirm the relation \eqref{VO-fO} for $V_3$ with the scalar function, 
\begin{align}
f = \frac{1}{m^2} \int E_\mu(\tau_1) \left[ R_{12} R_{23}  \partial^{(\mu} E^{\nu)}(\tau_2)  
+ \frac{1}{2} (\theta_{12} R_{23} + R_{12} \theta_{23})F^{\mu\nu}(\tau_2)  \right] E_\nu(\tau_3)\,.
\end{align}
The ``E-type" term has a Newtonian counterpart, while the ``F-type" term does not. %
The computation for $V_1$ and $V_2$ proceeds similarly 
and confirm that the same $f$ generates $V = V_1 + V_2 + V_3$. 

Repeating the same exercise for other topologies, we finally arrive at the full $\chi_{(3)}$:  
\begin{align}
\begin{split}
    \chi_{(3)}
    & \, = \frac{q^3}{m^2} \int \mathcal{I}_{(3)}(\tau_1,\tau_2,\tau_3) \,d\tau_1  d\tau_2  d\tau_3 \,,
    \\
    \mathcal{I}_{(3)} &= \frac{1}{3} R_{12} R_{23}  E_\nu(\tau_1)\partial^{(\nu} E^{\rho )}(\tau_2) E_\rho(\tau_3)    + \frac{1}{6}(\theta_{12}R_{23} + R_{12}\theta_{23})  E_\nu(\tau_1) F^{\nu \rho}(\tau_2) E_\rho(\tau_3) 
    \\
    & \hspace{0.3cm} +\frac{1}{12} R_{12} R_{13} \partial^{(\nu} E^{\rho)}(\tau_1) E_\nu(\tau_2) E_\rho(\tau_3) + \frac{1}{24}(R_{12}\theta_{13} -\theta_{12}R_{13}) F^{\nu \rho}(\tau_1) E_\nu(\tau_2) E_\rho(\tau_3)
    \\
    & \hspace{0.3cm} +\frac{1}{12} R_{13} R_{23} E_\nu(\tau_1) E_\rho(\tau_2) \partial^{(\nu} E^{\rho)}(\tau_3) + \frac{1}{24}(\theta_{13}R_{23} - R_{13}\theta_{23}) E_\nu(\tau_1) E_\rho(\tau_2) F^{\nu \rho}(\tau_3) \,.
\end{split}
\label{chi3-EM}
\end{align}

\paragraph{Comparison with WQFT} 

The extraction of $\chi_{(2)}$, $\chi_{(3)}$ above suggests how the procedure will generalize to higher orders. 
Before doing so, let us examine how the tree diagrams arising from the Magnus expansion match with those from the WQFT Feynman rules. 

It is straightforward to read off the WQFT vertex factors from the interaction term 
\begin{align}
\begin{split}
        S_\mathrm{int} &= q \int A_\mu dx^\mu 
        \\
        &= q \int A_\mu(b+v\tau +\delta x) (v^\mu + \delta{\dot{x}}^\mu) d\tau 
        \\
        &= q \int \left[ \mathcal{V}^{(0)} + \mathcal{V}^{(1)} +  \frac{1}{2!}  \mathcal{V}^{(2)} + \frac{1}{3!} \mathcal{V}^{(3)} + \cdots  \right]\,.
\end{split}
\end{align}
Throwing away some total derivative terms, we find 
\begin{align}
    \begin{split}
        \mathcal{V}^{(0)} &= v^\mu A_\mu \,,
        \\
        \mathcal{V}^{(1)} &=  E_\mu \delta x^\mu \,,
        \\ 
        \mathcal{V}^{(2)} &=  [\partial_{\mu_1} E_{\mu_2}] \delta x^{\mu_1} \delta x^{\mu_2} +  F_{\mu_1\mu_2} \delta x^{\mu_1} \delta \dot{x}^{\mu_2} \,,
        \\
        &\;\, \vdots 
        \\
        \mathcal{V}^{(n)} &= [\partial_{\mu_1} \cdots \partial_{\mu_{n-1}} E_{\mu_n}] \,\delta x^{\mu_1} \cdots \delta x^{\mu_n} 
        \\
        &\qquad + (n-1)\, [\partial_{\mu_1} \cdots \partial_{\mu_{n-2}} F_{\mu_{n-1} \mu_n} ]  \delta x^{\mu_1} \cdots \delta x^{\mu_{n-1}} \,\delta \dot{x}^{\mu_{n}} \,.
    \end{split}
    \label{vertex-factors}
\end{align}
From $\mathcal{V}^{(2)}$ and on, the vertex factor carries two terms at the same order. 
We call them ``E-type" and ``F-type". 
When applying the Feynman rules, we should consider all possible Wick contractions. It is useful to write the vertex factor in a form symmetric in all the indices. The symmetrization is manifest for the ``E-type" vertex factor in \eqref{vertex-factors}, 
but not for the ``F-type" vertex factor. 
Enforcing the symmetrization, we get an alternative expression for the F-type vertex factor, 
\begin{align}
    \mathcal{V}^{(n)}_F = \frac{1}{n} \sum_{k=1}^n \sum_{\ell \neq k} \left(\prod_{r\neq k, \ell} \partial_{\mu_r}\right) F_{\mu_\ell \mu_k} \left(\prod_{r\neq k, \ell} \delta x^{\mu_r}\right) \delta x^{\mu_\ell} \delta\dot{x}^{\mu_k} \,.
      \label{vertex-factors-3}
\end{align}

The WQFT uses these vertices and the propagators $\langle \delta x^\mu \delta x^\nu \rangle$, $\langle \delta x^\mu \delta \dot{x}^\nu \rangle$, $\langle \delta \dot{x}^\mu \delta \dot{x}^\nu \rangle$ to compute the eikonal. 
Clearly, $\chi_{(2)}$ in \eqref{chi2-EM} and $\chi_{(3)}$ in \eqref{chi3-EM} are fully consistent with the expectation from the WQFT. One extra information from the Magnus expansion is the causality prescription \cite{Kim:2024svw}.

\subsection{Eikonal to all orders}

We have seen how the eikonal can be extracted from the sum of nested commutators produced by the Magnus expansion, 
and how the results agree with the WQFT Feynman rules. In this subsection, we extend the extraction technique to all orders. 
For the E-type vertex factors, the computation is essentially the same as in the Newtonian case. More work is needed for the F-type vertex factors.

\paragraph{Vertex attached to external leg} 

Consider the vector field, 
\begin{align}
   Q_i =  \left( \prod_{k\in N(i)} \theta_{ik} D_k^{\mu_k} \right) E_\rho(\tau_i) D_i^\rho \,,
\end{align}
where $N(i)$ is the set of neighbors of the $i$-th vertex. 
Using some useful facts, 
\begin{align}
    D_k^\mu - D_i^\mu = \frac{\tau_{ik}}{m} \partial^\mu \,,
    \quad 
    D_i^\mu E_\rho(\tau_i) = \frac{1}{m} F_\rho{}^\mu(\tau_i) \,,
    \quad 
    D_i^\mu D_i^\nu E_\rho(\tau_i) = 0 \,,
    \label{DEF}
\end{align}
we can rewrite $Q_i$ as 
\begin{align}
\begin{split}
   Q_i &= Q_i^{(0)} + Q_i^{(1)}
      \\
   &= \left( \prod_{k\in N(i)} \frac{R_{ik}}{m}  \partial^{\mu_k} \right) E_\rho(\tau_i) D_i^\rho 
    + \sum_{k\in N(i)} \theta_{ik} D_i^{\mu_k} \left( \prod_{\ell \in N(i)\backslash k} \frac{R_{i\ell}}{m}  \partial^{\mu_\ell} \right) E_\rho(\tau_i) D_i^\rho 
     \,.  
\end{split}
\end{align}
We are going to show that the E-type vertex factor comes entirely from $Q_i^{(0)}$, 
whereas the F-type vertex factor receives contributions from both $Q_i^{(0)}$ and $Q_i^{(1)}$. 

Let $n_i = |N(i)|$ be the valence (number of neighbors) of the $i$-th vertex. 
To specify a neighbor one at a time while maintaining the symmetry among them, we write 
\begin{align}
  Q_i^{(0)} =  \left( \prod_{k\in N(i)} \frac{R_{ik}}{m} \right) \frac{1}{n_i} \sum_{k\in N(i)}  \partial^{\mu_k} \left( \prod_{\ell \in N(i)\backslash k}  \partial^{\mu_\ell} \right) E_\rho(\tau_i) D_i^\rho \,.
\end{align}
Now we apply the Bianchi identity \eqref{Bianchi-E} to $\partial^{\mu_k} E_\rho(\tau_i)$:
\begin{align}
    \partial^{\mu_k} E_\rho(\tau_i) = \partial_\rho E^{\mu_k}(\tau_i) + \frac{d F_{\rho}{}^{\mu_k}(\tau_i)}{d\tau_i} \,.
\end{align}
From the $\partial_\rho E^{\mu_k}(\tau_i)$ term, we get 
\begin{align}
\begin{split}
    Q_i^{(0,a)} &=\left( \prod_{k\in N(i)} \frac{R_{ik}}{m} \right)   \partial_\rho \left(  \frac{1}{n_i} \sum_{k\in N(i)}  \left( \prod_{\ell \in N(i)\backslash k}  \partial^{\mu_\ell} \right) E^{\mu_k}(\tau_i)  \right) D_i^\rho  = (\partial_\rho W_{E,i}) D_i^\rho \,,
    \\
  &\qquad W_{E,i}  = \left(\prod_{k\in N(i)} \frac{R_{ik}}{m} \right) \partial^{(\mu_1} \cdots \partial^{\mu_{n_i-1}} E^{\mu_{n_i})} (\tau_i)\,.
\end{split}
\label{WEO}
\end{align}
For the $d F_{\rho}{}^{\mu_k}(\tau_i)/d\tau_i$ term, we perform an integration by parts (IBP). The time-derivative acts either on the external $D_i^\rho$, or on the product of $R_{ik}$'s.  We can write the former contribution as, using \eqref{DEF}, 
\begin{align}
    Q_i^{(0,b)} &= - (D^{\rho}_i W_{E,i}) \partial_\rho \,. 
\end{align}
Adding $Q_i^{(0,a)}$ and $Q_i^{(0,b)}$, while leaving aside the remainder to be called $Q_i^{(0,c)}$, we find 
\begin{align}
  \left(  Q_i^{(0,a)} + Q_i^{(0,b)} \right) O = \{ W_{E,i} , O \}_i \,, 
\end{align}
where we used \eqref{Poisson-D-shift} with $\tau=\tau_i$. 
Thus, we successfully extracted the E-type vertex factor (contracted with neighboring vertices). 

Our next goal is to extract the F-type vertex factor:
\begin{align}
 \left(  Q_i^{(0,c)} + Q_i^{(1)}  \right) O =   \{ W_{F,i} , O \}_i = (\partial_\rho W_{F,i}) D_i^\rho O \,.
 \label{WFO}
\end{align}
Here, we dropped $D_i^\rho W_{F,i} = 0$ which results from $D_i^\rho F_{\mu\nu}(\tau_i) = 0$. 
To see how the F-type arises, we collect the contributions.
\begin{align}
\begin{split}
   Q^{(0,c)}_i &= -\frac{1}{n_i} \sum_k \frac{\theta_{ik}}{m} \left( \prod_{\ell \neq k} \frac{R_{i\ell}}{m} \partial^{\mu_\ell} \right) F_\rho{}^{\mu_k}
   D^\rho 
   \\
   &\qquad 
   - \frac{1}{n_i} \sum_{k\neq \ell} \frac{R_{ik}}{m} \frac{\theta_{i\ell}}{m} \left( \prod_{r \neq k,\ell} \frac{R_{ir}}{m} \partial^{\mu_q} \right) \partial^{\mu_\ell} F_\rho{}^{\mu_k}  D^\rho\,,
   \\
    Q^{(1)}_i &= \sum_k \frac{\theta_{ik}}{m} \left( \prod_{\ell \neq k} \frac{R_{i\ell}}{m} \partial^{\mu_\ell} \right) F_\rho{}^{\mu_k} D^\rho \,.
\end{split}
\end{align}
It is understood that an IBP has been performed on $Q^{(0,c)}$. It is natural to add up the first term of $Q^{(0,c)}$ and $Q^{(1)}$. 
Then we rewrite the sum slightly to combine it with the second term of $Q^{(0,c)}$. The intermediate steps are
\begin{align}
\begin{split}
   Q^{(0,c)}_i+   Q^{(1)}_i &= \frac{n_i-1}{n_i} \sum_k \frac{\theta_{ik}}{m} \left( \prod_{\ell \neq k} \frac{R_{i\ell}}{m} \partial^{\mu_\ell} \right)  F_\rho{}^{\mu_k}
   D^\rho 
   \\
   &\qquad 
   - \frac{1}{n_i} \sum_{k\neq \ell} \frac{R_{ik}}{m} \frac{\theta_{i\ell}}{m} \left( \prod_{r \neq k,\ell} \frac{R_{ir}}{m} \partial^{\mu_r} \right) \partial^{\mu_\ell} F_\rho{}^{\mu_k}  D^\rho
\\
&=\frac{1}{n_i} \sum_{k\neq \ell} \frac{(\theta_{ik} R_{i\ell} -  R_{ik} \theta_{i\ell})}{m^2} \left( \prod_{r \neq k,\ell} \frac{R_{ir}}{m} \partial^{\mu_r} \right) \partial^{\mu_\ell}F_\rho{}^{\mu_k}  D^\rho \,.
\end{split}
\end{align}
Exchanging the labels $k$ and $\ell$ for the second term, we get 
\begin{align}
\begin{split}
   Q^{(0,c)}_i + Q^{(1)}_i &= \frac{1}{n_i} \sum_{k\neq \ell} \frac{\theta_{ik} R_{i\ell}}{m^2} \left( \prod_{r \neq k,\ell} \frac{R_{ir}}{m} \partial^{\mu_r} \right) (\partial^{\mu_\ell} F_\rho{}^{\mu_k} -\partial^{\mu_k} F_\rho{}^{\mu_\ell} ) D^\rho \,.
\end{split}
\end{align}
Applying the Bianchi identity once again, 
\begin{align}
   \partial^{\mu_\ell} F_\rho{}^{\mu_k} -\partial^{\mu_k} F_\rho{}^{\mu_\ell} = \partial_\rho F^{\mu_\ell \mu_k}   \,,
\end{align}
we pull out an overall $\partial_\rho$ as needed in \eqref{WFO}. The resulting (contracted) vertex factor $W_{F,i}$ agrees perfectly with the Feynman rule \eqref{vertex-factors-3}. 

\paragraph{Vertex not attached to external leg}

Consider the case where the vertex $i$ has $(n-1)$ incoming legs and one outgoing leg attached to another vertex, say, vertex $k$. 
Define the vector field $S_i$ as
\begin{align}
    S_i E_\rho(\tau_k) = \prod_{\ell \neq k} \left(\theta_{i\ell} D_\ell^{\mu_{\ell}}  \right) E_\nu(\tau_i) \theta_{ik} D_i^\nu E_\rho(\tau_k) \,.
\end{align}
We wish to show that, much like $Q_i$ we studied earlier, $S_i$ reproduces the E-type and F-type vertex factors. Again, we split $S_i$ into two pieces. 
\begin{align}
\begin{split}
   S^{(0)}_i = \prod_{\ell \neq k} \left(\frac{R_{i\ell}}{m} \partial^{\mu_{\ell}}  \right) E^\nu(\tau_i) \theta_{ik} D_{i,\nu}   \,, 
   \quad
   S^{(1)}_i = \sum_{\ell\neq k} \frac{\theta_{i\ell}}{m} \prod_{r \neq k,\ell} \left(\frac{R_{ir}}{m} \partial^{\mu_{r}}  \right) F^{\nu \mu_\ell}(\tau_i) D_{i,\nu} ,.
\end{split}
\end{align}
From $S_i^{(0)}$, we first pull out the part totally symmetric in $\{\mu_\ell\}$ and $\nu$ indices:
\begin{align}
    S_i^{(0,a)} = \frac{1}{n} \left[ \prod_{\ell\neq k}  \left(\frac{R_{i\ell}}{m} \partial^{\mu_\ell}  \right) E^\nu + \sum_{\ell\neq k} R_{i\ell} \prod_{r\neq k,\ell} \left(\frac{R_{ir}}{m} \partial^{\mu_{r}}  \right) \partial^\nu E^{\mu_\ell} \right] \,.
\end{align}
It gives the E-type vertex factor. 
The remaining part of $S_i^{(0)}$ gives 
\begin{align}
\begin{split}
   &\frac{n-1}{n} \prod_{\ell \neq k} \left(\frac{R_{i\ell}}{m} \partial^{\mu_\ell} \right) E^\nu - \frac{1}{n} \sum_{\ell\neq k} \frac{R_{i\ell}}{m} \prod_{r\neq k,\ell} \left(\frac{R_{ir}}{m} \partial^{\mu_{r}}  \right) \partial^\nu E^{\mu_\ell} D_\nu
     \\
     &= \frac{1}{n} \sum_{\ell} \frac{R_{i\ell}}{m} \prod_{r } \left(\frac{R_{ir}}{m} \partial^{\mu_{r}}  \right) (\partial^{\mu_\ell} E^\nu - \partial^\nu E^{\mu_\ell} ) D_\nu
    \\
    &= \frac{1}{n}  \sum_{\ell} \frac{R_{i\ell}}{m} \prod_{r} \left(\frac{R_{ir}}{m} \partial^{\mu_{r}}  \right) \left( \frac{d}{d\tau_i} F^{\nu \mu_\ell} \right) D_\nu
    \\
    &= -\frac{1}{n} \sum_{\ell}  \frac{d}{d\tau_i} \left(\frac{R_{i\ell}}{m}\prod_{r} \left(\frac{R_{ir}}{m} \partial^{\mu_{r}}  \right) \right) F^{\nu \mu_\ell} D_\nu 
    \\
    &\qquad 
    -\frac{1}{n} \sum_\ell  \frac{R_{i\ell}}{m}\prod_{r} \left(\frac{R_{ir}}{m} \partial^{\mu_{r}}  \right)  F^{\nu \mu_\ell}  \frac{d}{d\tau_i}\left(D_\nu \right) 
    \\
    &\equiv  S_i^{(0,b)} +   S_i^{(0,c)}  \,.
\end{split}
\label{E-leftover}
\end{align}
Between the second and third lines, we applied the Bianchi identity. Between the third and fourth, we did an IBP. The sum of $S_i^{(0,b)}$ and $S_i^{(1)}$ gives 
\begin{align}
\begin{split}
   S_i^{(0,b)} + S_i^{(1)} 
   &= 
   \frac{n-1}{n} \sum_{\ell \neq k} \frac{\theta_{i\ell}}{m}\prod_{r \neq k, \ell} \left(\frac{R_{ir}}{m} \partial^{\mu_{r}}  \right) F^{\nu\mu_\ell} D_\nu
   \\
   &\qquad - \frac{1}{n} \sum_{\ell\neq k} \frac{R_{i\ell}}{m} \frac{d}{d\tau_i} \left(\prod_{r\neq k, \ell} \frac{R_{ir}}{m} \partial^{\mu_{r}}  \right) F^{\nu\mu_\ell} D_\nu
   \\
   &= \frac{1}{n} \sum_{\ell,r} \frac{(\theta_{i\ell} R_{ir} - R_{i\ell} \theta_{ir} )}{m^2} \prod_{s\neq k, \ell, r} \left( \frac{R_{is}}{m} \partial^{\mu_s}\right) \partial^{\mu_r} F^{\nu \mu_\ell} D_\nu
   \\
   &\qquad+  \frac{1}{n} \sum_{\ell} \frac{\theta_{i\ell}}{m}\prod_{r} \left(\frac{R_{ir}}{m} \partial^{\mu_{r}}  \right) F^{\nu\mu_\ell} D_\nu \,, 
\end{split}
\end{align}
where we used $n-1 = (n-2) + 1$. 
Exchanging $\ell$ and $r$ labels in the $(-R_{i\ell} \theta_{ir})$ part of the sum and then applying the Bianchi identity, we get 
\begin{align}
\begin{split}
  S_i^{(0,b)} + S_i^{(1)} 
   &= \frac{1}{n} \sum_{\ell,r} \frac{\theta_{i\ell} R_{ir}}{m^2} \prod_{s} \left( \frac{R_{is}}{m} \partial^{\mu_s}\right) \partial^{\nu} F^{ \mu_r\mu_\ell} D_\nu
   \\
   &\qquad+  \frac{1}{n} \sum_{\ell} \frac{\theta_{i\ell}}{m}\prod_{r} \left(\frac{R_{ir}}{m} \partial^{\mu_{r}}  \right) F^{\nu\mu_\ell} D_\nu \,.   
\end{split}
\end{align}
These terms match the F-type vertex factor in \eqref{vertex-factors-3}, except that the $\langle \delta \dot{x} \delta x\rangle$ propagator along the $(ik)$ leg is missing. But, the $S^{(0,c)}$ term we left behind in \eqref{E-leftover} precisely accounts for the missing piece, 
so the sum $S_i^{(0,b)} + S_i^{(0,c)} + S_i^{(1)}$ reproduces the F-type vertex factor.

\section{Gravitational background} \label{sec:Grav-background}

\subsection{Eikonal to all orders}

In a Hamiltonian picture, the point particle action in a background metric reads  
\begin{align}
    S = \int \left[p_\mu \dot{x}^\mu - \frac{\kappa}{2}\left( g^{\mu\nu}(x) p_\mu p_\nu + m^2 \right) \right]d\tau \,.
    \label{g-particle-action}
\end{align}
The variation of the action gives (with $\kappa=1/m$) 
\begin{align}
\begin{split}
    \frac{dx^\mu}{d\tau} &= \frac{1}{m} g^{\mu\nu}(x) p_\nu \,,
    \\
    \frac{dp_\mu}{d\tau} &= -\frac{1}{2m} \partial_\mu g^{\rho\sigma}(x) p_\rho p_\sigma \,.
\end{split}
\label{EOM-gravity}
\end{align}
The $p$-EOM does not look covariant, but it can be modified to a covariant form, 
\begin{align}
    \frac{Dp_\mu}{d\tau} :=    \frac{dp_\mu}{d\tau} - \frac{dx^\rho}{d\tau} \Gamma^\sigma_{\rho \mu} p_\sigma = 0 \,, 
\end{align}
recovering the geodesic equation for the particle. 

In a perturbation theory, we treat $g^{\mu\nu}(x)$ as a tensor field in special relativity. Noting that the RHS of the EOM \eqref{EOM-gravity} only involves the \emph{inverse} metric $g^{\mu\nu}(x)$,  
we find it convenient to work with the metric deviation $\tilde{h}^{\mu\nu}$ defined as 
\begin{align}
    g^{\mu\nu} = \eta^{\mu\nu} - \tilde{h}^{\mu\nu} 
    \quad 
    \mbox{as opposed to} 
    \quad 
    g_{\mu\nu} = \eta_{\mu\nu} + h_{\mu\nu} \,.
\end{align}
The two fields $h_{\mu\nu}$ and $\tilde{h}^{\mu\nu}$ are related by 
(indices are raised and lowered by the \emph{flat} metric) 
\begin{align}
  \tilde{h}^{\mu\nu} =  h^{\mu\nu} - h^{\mu\rho} h_\rho{}^\nu + \mathcal{O}(h^3) \,.
  \label{h-tilde-vs-h}
\end{align}
In what follows, by definition, the $n$-th order eikonal $\tilde{\chi}_{(n)}$ should be proportional to the $n$-th power of $\tilde{h}^{\mu\nu}$ and its derivatives. 

\paragraph{Interaction picture}

For the CIP, we use the metric fluctuation $\tilde{h}^{\mu\nu}$ along the background trajectory $\bar{x}^\mu(\tau) = b^\mu + v^\mu \tau$. 
We also introduce scalar and vector quantities:
\begin{align}
    \tilde{h}(\tau) \equiv \frac{1}{2} \tilde{h}^{\mu\nu}(\bar{x}(\tau)) v_\mu v_\nu \,,
    \quad \tilde{h}^\mu(\tau) \equiv \tilde{h}^{\mu\nu}(\bar{x}(\tau)) v_\nu \,, 
    \qquad 
    v_\mu \equiv \frac{p_\mu}{m}\,.
    \label{h-short}
\end{align}
The action \eqref{g-particle-action} (unlike in the EM background) admits a straightforward Hamiltonian deformation with $H_I = -m \tilde{h}(x+v\tau)$.
We can apply the Magnus expansion immediately. 
At the 1st order, 
\begin{align}
        \tilde{\chi}_{(1)} = m \int \tilde{h}(b+v\tau)  d\tau \,.
        \label{chi1-g}
\end{align}
The 2nd order expansion gives 
\begin{align}
\begin{split}
    \tilde{\chi}_{(2)} &= - \frac{1}{2} \int \theta_{12} \{ V(\tau_1), V(\tau_2) \} d\tau_1 d\tau_2 
\\
    &= \frac{m}{2} \int R_{12} \partial_\nu \tilde{h}(\tau_1) \partial^\nu \tilde{h}(\tau_2) + \frac{m}{2} \int \theta_{12} \left[ \tilde{h}_\nu(\tau_1) \partial^\nu \tilde{h}(\tau_2) - \partial_\nu \tilde{h}(\tau_1) \tilde{h}^\nu(\tau_2) \right] \,. 
\end{split}
    \label{chi2-g}
\end{align}
The ``non-Newtonian" second term is due to the explicit $v$-dependence of the scalar $\tilde{h}$. 

We can state a simple rule to produce all non-Newtonian terms 
as we proceed to higher orders. We note two facts. One is about the Poisson bracket,
\begin{align} \label{Poisson_general}
    \{ f(\tau_i), g(\tau_j) \} = -\frac{\tau_{ij}}{m} \partial_\mu f(\tau_i) \partial^\mu g(\tau_j) + \partial_\mu f(\tau_i) D^\mu g(\tau_j) - D_\mu f(\tau_i) \partial^\mu g(\tau_j) \,,
\end{align}
The other is the relation between the scalar, vector, tensor quantities:
\begin{align}
    D_\mu \tilde{h} = \frac{1}{m} \tilde{h}_\mu \,, 
    \quad 
    D_\mu \tilde{h}_\nu = \frac{1}{m} \tilde{h}_{\mu\nu} \,, \quad 
    D_\mu D_\nu D_\sigma \tilde{h} = 0 \,.
\end{align}
While computing the eikonal, the first term in the RHS of \eqref{Poisson_general} gives the Newtonian term, 
while the other two terms generate the non-Newtonian terms. 
The two facts above suggest that we can compute $\tilde{\chi}_{(n)}$ by starting with the Newtonian eikonal and sum over all possible substitutions, 
\begin{align}
    \frac{1}{m}R_{ij} \frac{\partial}{\partial x_i^\mu} \frac{\partial}{\partial x_{j,\mu}} \quad \rightarrow \quad \theta_{ij} \left( D^\mu_i \frac{\partial}{\partial x_j^\mu}  - \frac{\partial}{\partial x_i^\mu} D^\mu_j \right) \,.
\end{align}

\subsection{WQFT vs. Magnus}

We begin with the interaction part of the action, 
\begin{align}
    S_\mathrm{int} = \frac{m}{2} \int h_{\mu\nu}(x) \dot{x}^\mu \dot{x}^\nu d\tau \,. 
    \label{int-Lagrangian}
\end{align}
Expanding it with $x = b + vt + \delta x$, 
we get
\begin{align}
\begin{split}
\frac{1}{m} \mathcal{V}^{(n)} &= \frac{1}{2}  (\partial_{\mu_1} \cdots \partial_{\mu_n} h_{\rho \sigma} v^\rho v^\sigma )
\delta x^{\mu_1} \cdots \delta x^{\mu_n} 
\\
&\quad +  n (\partial_{\mu_1} \cdots \partial_{\mu_{n-1}} h_{\mu_n \sigma} v^\sigma) \delta x^{\mu_1} \cdots \delta x^{\mu_{n-1}} \delta \dot{x}^{\mu_n} 
\\
&\quad +   \frac{n(n-1)}{2} (\partial_{\mu_1} \cdots \partial_{\mu_{n-2}} h_{\mu_{n-1} \mu_n} ) \delta x^{\mu_1} \cdots \delta x^{\mu_{n-2}} \delta \dot{x}^{\mu_{n-1}} \delta \dot{x}^{\mu_n} \,.
\end{split}
\label{gravity-vf-naive}
\end{align}
General covariance is far from manifest in this approach. 
Even if the vertex factors are written in terms of covariant quantities such as $\delta\Gamma(g,\eta)$, $R(g)$, 
if they rely on the straight line trajectory, $\bar{x}^\mu(\tau) = b^\mu + v^\mu\tau$, 
they are not really covariant. 
But since the underlying theory has general covariance, the observables should be unambiguous.

At each order, we have three types of terms.
Schematically, 
\begin{align}
    \frac{1}{m} \mathcal{V}^{(n)} = (\partial^n h ) (\delta x)^n + (\partial^{n-1} h) (\delta x)^{n-1} (\delta\dot{x})+ (\partial^{n-2} h) (\delta x)^{n-2} (\delta\dot{x})^{2} \,.
    \label{EFG-V}
\end{align}
In analogy with the E/F-type electromagnetic vertex factors from the previous section, we name them E/F/G-type vertex factors. IBPs can transform an E-type term to an F-type term, etc. So, the separation into three types is somewhat ambiguous.

\paragraph{Linearized covariance} 
The vertex factors are defined up to total derivatives. 
Using IBP, we may write the vertex factors in a superficially covariant form. 
For example, we can rewrite the first three vertex factors as 
\begin{align}
    \begin{split}
        \frac{1}{m} \mathcal{V}^{(0)} &= \frac{1}{2} h_{\mu\nu} v^\mu v^\nu \,,
        \\
         \frac{1}{m}  \mathcal{V}^{(1)} &= \frac{1}{2} (\partial_\rho h_{\mu\nu} - \partial_\mu h_{\nu\rho} - \partial_\nu h_{\mu\rho} ) v^\mu v^\nu \delta x^\rho \,,
          \\
          \frac{1}{m} \mathcal{V}^{(2)} &=  \frac{1}{2}\left( \partial_\rho \partial_\sigma h_{\mu\nu} + \partial_\mu \partial_\nu h_{\rho\sigma} - \partial_\rho \partial_\nu h_{\mu\sigma} - \partial_\mu \partial_\sigma h_{\rho\nu} \right) v^\mu v^\nu \delta x^\rho \delta x^\sigma 
          \\
          &\quad  +   (\partial_\mu h_{\rho\sigma} + \partial_\rho h_{\mu\sigma} - \partial_\sigma h_{\mu\rho})v^\mu \delta x^\rho \delta\dot{x}^\sigma +  h_{\rho\sigma} \delta\dot{x}^\rho\delta \dot{x}^\sigma 
          \,.
        \end{split}
        \label{EFG-0123}
\end{align}
We recognize the \emph{linearized} Levi-Civita connection and Riemann curvature:
\begin{align}
\begin{split}
    \bar{\Gamma}_{\rho \mu \nu} &\equiv \frac{1}{2}( \partial_\mu h_{\nu\rho} + \partial_\nu h_{\mu\rho} - \partial_\rho h_{\mu\nu} ) \,,
    \\
    \bar{R}_{\mu \rho \nu \sigma} & \equiv \partial_\nu \bar{\Gamma}_{\mu \sigma \rho} - \partial_\sigma \bar{\Gamma}_{\mu \nu \rho} 
    \\
    &=\frac{1}{2}\left(   \partial_\rho \partial_\nu h_{\mu\sigma} + \partial_\mu \partial_\sigma h_{\rho\nu}  - \partial_\mu \partial_\nu h_{\rho\sigma} - \partial_\rho \partial_\sigma h_{\mu\nu} \right) \,.
    \label{Gamma-R-linear}
\end{split}
\end{align}
In terms of these tensors, the vertex factors read 
\begin{align}
    \begin{split}
      \frac{1}{m}  \mathcal{V}^{(0)} &= \frac{1}{2} h_{\mu\nu} v^\mu v^\nu \,,
        \\
        \frac{1}{m}  \mathcal{V}^{(1)} &= - \bar{\Gamma}_{\rho \mu\nu} v^\mu v^\nu \delta x^\rho \,,
          \\
         \frac{1}{m} \mathcal{V}^{(2)} &=   -\bar{R}_{\mu\rho\nu\sigma} v^\mu v^\nu \delta x^\rho \delta x^\sigma 
          + 2 \bar{\Gamma}_{\sigma \rho\mu} v^\mu \delta x^\rho \delta\dot{x}^\sigma 
           + h_{\rho\sigma} \delta\dot{x}^\rho\delta \dot{x}^\sigma 
          \,.
        \end{split}
        \label{EFG-0123b}
\end{align}
We can continue to derive the linear-level-covariant expressions to all orders. 
One way to write the vertex factors is
\begin{align}
\begin{split}
    \frac{1}{m} \mathcal{V}^{(n)} & = -[\partial_{\mu_1} \cdots \partial_{\mu_{n-2}} \bar{E}_{\mu_{n-1}\mu_n} ] \delta x^{\mu_1} \cdots \delta x^{\mu_n}
    \\
    &\quad +2 (n-1)   [\partial_{\mu_1} \cdots \partial_{\mu_{n-2}} \bar{\Gamma}_{\mu_{n} \mu_{n-1} \sigma} v^\sigma] \delta x^{\mu_1} \cdots \delta x^{\mu_{n-1}} \delta \dot{x}^{\mu_n}
    \\
    &\quad -(n-2)  [\partial_{\mu_1} \cdots \partial_{\mu_{n-3}} \partial_\sigma 
 \bar{\Gamma}_{\mu_n \mu_{n-1} \mu_{n-2}} v^\sigma]  \delta x^{\mu_1} \cdots  \delta x^{\mu_{n-1}} \delta \dot{x}^{\mu_n}
    \\
    &\quad + \frac{n(n-1)}{2}   [\partial_{\mu_1} \cdots \partial_{\mu_{n-2}} h_{\mu_{n-1} \mu_{n}}] \delta x^{\mu_1} \cdots \delta x^{\mu_{n-2}} \delta \dot{x}^{\mu_{n-1}} \delta \dot{x}^{\mu_n}
   \,,
\end{split}
\label{gravity-vf-a}
\end{align}
where we introduced the ``tidal tensor":
\begin{align}
\bar{E}_{\rho\sigma} =  \bar{R}_{\mu \rho \nu \sigma} v^\mu v^\nu  \,.
\end{align}
Using an identity, where some terms cancel due to (anti-)symmetry, 
\begin{align}
\begin{split}
    & \bar{\Gamma}_{\mu_n \mu_{n-2} \mu_{n-1}}\delta x^{\mu_{n-2}} \delta \dot{x}^{\mu_{n-1}} \delta \dot{x}^{\mu_n}
    \\
    & = \frac{1}{2} [\partial_{\mu_{n-2}}h_{\mu_{n-1} \mu_{n}} +\partial_{\mu_{n-1}}h_{\mu_{n-2} \mu_{n}} -\partial_{\mu_{n}}h_{\mu_{n-2} \mu_{n-1}}] \delta x^{\mu_{n-2}} \delta \dot{x}^{\mu_{n-1}} \delta \dot{x}^{\mu_n}
    \\
    & = \frac{1}{2}  \partial_{\mu_{n-2}}h_{\mu_{n-1} \mu_{n}} \delta x^{\mu_{n-2}} \delta \dot{x}^{\mu_{n-1}} \delta \dot{x}^{\mu_n} \,,
\end{split}
\label{h-to-Gamma}
\end{align}
we can remove $h$ in favor of $\bar{\Gamma}$ and remove $\bar{\Gamma}$ in favor of $\bar{R}$ using \eqref{Gamma-R-linear}. The result is
\begin{align}
\begin{split}
    \frac{1}{m} \mathcal{V}^{(n)} & = -[\partial_{\mu_1} \cdots \partial_{\mu_{n-2}} \bar{E}_{\mu_{n-1}\mu_n} ] \delta x^{\mu_1} \cdots \delta x^{\mu_n}
    \\
    &\quad +2 (n-1)   [\partial_{\mu_1} \cdots \partial_{\mu_{n-3}} \bar{R}_{\mu_{n} \mu_{n-1} \mu_{n-2} \sigma} v^\sigma] \delta x^{\mu_1} \cdots \delta x^{\mu_{n-1}} \delta \dot{x}^{\mu_n}
    \\
    &\quad +n  [\partial_{\mu_1} \cdots \partial_{\mu_{n-3}} \partial_\sigma 
 \bar{\Gamma}_{\mu_n \mu_{n-1} \mu_{n-2}} v^\sigma]  \delta x^{\mu_1} \cdots  \delta x^{\mu_{n-1}} \delta \dot{x}^{\mu_n}
    \\
    &\quad + n(n-1)   [\partial_{\mu_1} \cdots \partial_{\mu_{n-3}} \bar{\Gamma}_{ \mu_{n}\mu_{n-1}\mu_{n-2}}] \delta x^{\mu_1} \cdots \delta x^{\mu_{n-2}} \delta \dot{x}^{\mu_{n-1}} \delta \dot{x}^{\mu_n}
   \,.
\end{split}
\label{gravity-vf-b}
\end{align}
For a WQFT computation in a flat Minkowski background, any representation, \eqref{gravity-vf-naive} or \eqref{gravity-vf-a} or \eqref{gravity-vf-b}, 
should be as good as any other.

\paragraph{WQFT vs. Magnus} 

In the Lagrangian picture, we can use the WQFT vertex factors (based on $h_{\mu\nu}$), supplemented with the causality prescription, to compute the eikonal $\chi_{(n)}$. 
In the Hamiltonian picture, we can run the Magnus expansion with the ``scalar potential" $-m\tilde{h}$ (based on $\tilde{h}^{\mu\nu}$) to compute the eikonal $\tilde{\chi}_{(n)}$.

The classical limit of the S-matrix equivalence theorem \cite{Chisholm:1961tha,Kamefuchi:1961sb} 
guarantees that $\chi$ and $\tilde{\chi}$ should be equal in content. 
But the equivalence does not hold order-by-order. The non-linear relation between $h$ and $\tilde{h}$ induces mixing among different orders. 
From a perturbative point of view, the expected relation is  
\begin{align}
        \sum_{n=1}^m \chi_{(n)} = \sum_{n=1}^m \tilde{\chi}_{(n)} + \mathcal{O}(h^{m+1}) \,.
        \label{non-linear-equivalence}
\end{align}
On the Hamiltonian/Magnus side, each term is proportional to a product of $R_{ij}$ and $\theta_{ij}$. 
On the Lagrangian/WQFT side, the products often carry $\delta_{ij} = \delta(\tau_i -\tau_j) \propto \langle \delta\dot{x}(\tau_1) \delta\dot{x}(\tau_2)\rangle$ in addition to $R_{ij}$ and $\theta_{ij}$. 
The $\delta_{ij}$ factor turns an $n$-point tree to an $(n-1)$-point tree. Such a ``contraction" is needed 
to establish the non-linear equivalence \eqref{non-linear-equivalence}.

Recall that $\tilde{\chi}_{(n)}$ 
is a degree $n$ functional of $\tilde{h}_{\mu\nu}$. 
We can express $\tilde{\chi}_{(n)}$ 
as a functional of $h_{\mu\nu}$  by the substitution \eqref{h-tilde-vs-h}.
Let $\tilde{\chi}_{(n)k}$ be the sum of degree $(n+k)$ terms in $h_{\mu\nu}$. 
Working out the Magnus expansion and comparing the result with the WQFT side, we observe that $\tilde{\chi}_{(n)0}$ agrees with 
the subset of $\chi_{(n)}$ where all diagrams containing one or more $\delta_{ij}$ propagators are discarded. 
Then, we expect that $\tilde{\chi}_{(n)k}$ should match 
the subset of $\chi_{(n+k)}$ where precisely $k$ of the propagators are of the $\delta_{ij}$ type.

\begin{figure}[htbp]
    \centering
    \includegraphics[width=0.42\linewidth]{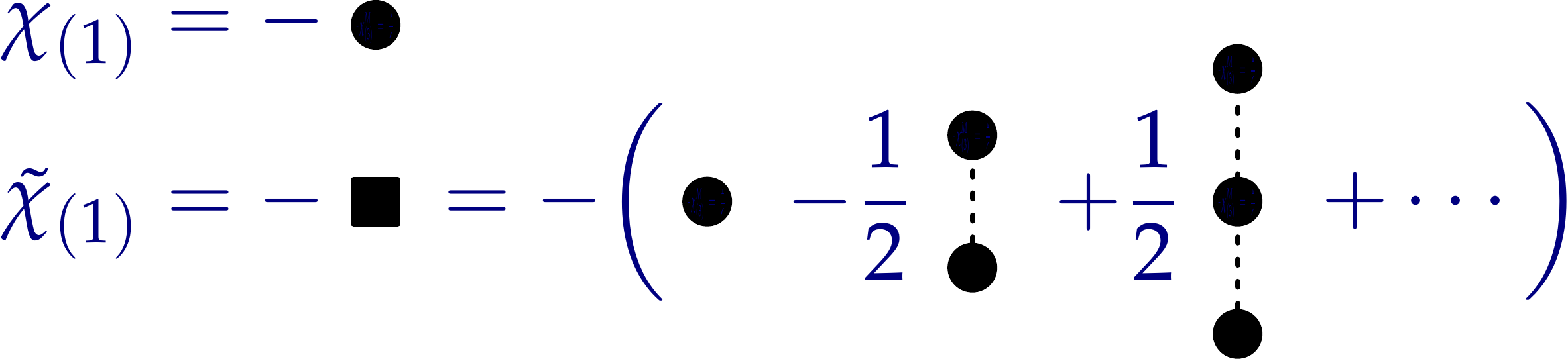}
    \caption{WQFT vs. Magnus at the leading order. We align the $\delta_{ij}$ propagators vertically to emphasize the absence of time-ordering.}
    \label{fig:chi1-g-comparison}
\end{figure}

To get a glimpse of the matching, we begin with linear trees. 
At the 1st order, recall 
\begin{align}
        \tilde{\chi}_{(1)} = m \int \tilde{h}(b+v\tau)  d\tau 
\quad \mbox{vs.} \quad 
    \chi_{(1)} = m \int h(b+v\tau) d\tau \,.
\end{align}
Expanding $\tilde{\chi}_{(1)}$ in $h$, we get 
\begin{align}
  \frac{1}{m}\left(\tilde{\chi}_{(1)} - \chi_{(1)} \right) = \frac{1}{2} 
  \int ( - h_\mu h^\mu +  h_\mu h^{\nu\nu} h_\nu- h_\mu h^{\mu \nu} h_{\nu\rho} h^\rho + 
  \cdots )\,. 
  \label{full-contraction-linear} 
\end{align}
The comparison is depicted in Figure~\ref{fig:chi1-g-comparison}. 
The $n$-th order term on the RHS should come from 
the ``fully contracted" term of $\chi_{(n)}$. 
It is easy to see that only linear trees can contribute. 
The WQFT vertices are at most quadratic in $\delta \dot{x}$. So, trees with valence 3 or higher cannot be fully contracted. 
Note that the coefficients $(1/2)(-1)^{n-1}$ in \eqref{full-contraction-linear} are  precisely 
the Feynman symmetry factors of linear trees in the WQFT. They are compatible with the Magnus expansion through 
the sum rules discussed in \cite{Kim:2024svw}. 

\begin{figure}[htbp]
    \centering
    \includegraphics[width=0.48\linewidth]{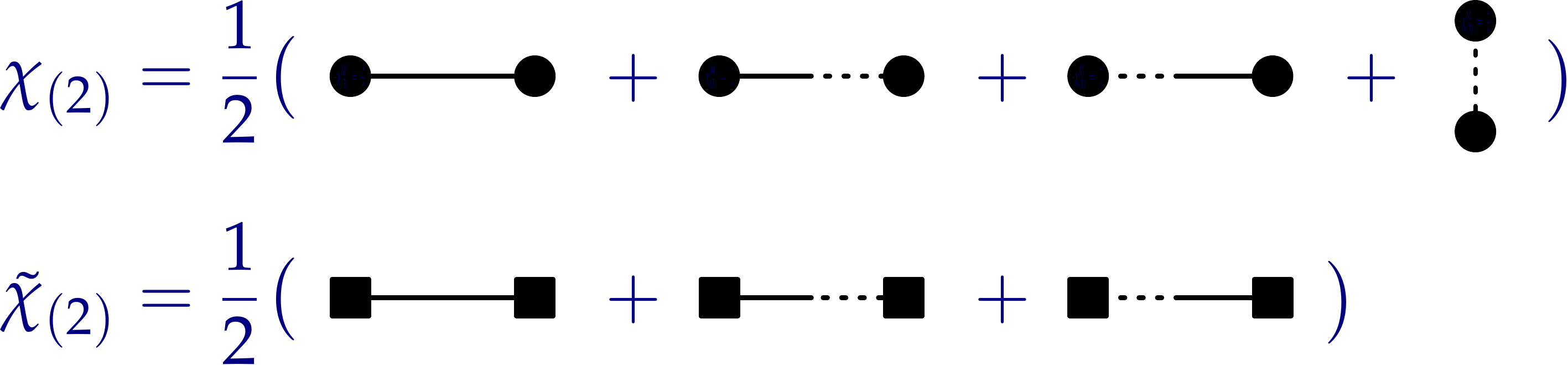}
    \caption{WQFT vs. Magnus at the 2nd order.}
    \label{fig:chi2-g-comparison}
\end{figure}

For the first example of the contraction, consider $\tilde{\chi}_{(2)}$ and $\chi_{(2)}$ as depicted in Figure~\ref{fig:chi2-g-comparison}. The only contraction term in $\chi_{(2)}$ is identified with $\tilde{\chi}_{(1)1}$ in Figure~\ref{fig:chi1-g-comparison}. 
$\tilde{\chi}_{(2)0}$ is equal to $\chi_{(2)}$ minus the  contraction term. 

\begin{figure}[htbp]
    \centering
    \includegraphics[width=0.46\linewidth]{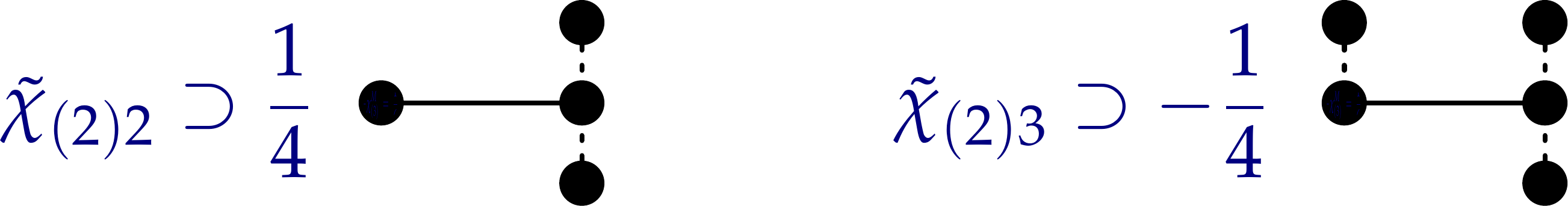}
    \caption{Examples of non-linear contracted trees.}
    \label{fig:chi2k}
\end{figure}

In $\tilde{\chi}_{(2)k}$ with $k\ge 1$, 
many terms correspond to linear trees in $\chi_{(n+k)}$. 
The first few examples of \emph{non-linear} trees include those in Figure~\ref{fig:chi2k}. 
For the first example, $\tilde{\chi}_{(2)2}$, 
there is only one oriented tree in $\chi_{(4)}$ with a non-vanishing $\omega$ coefficient: $\omega(Y_4) = -1/12$ 
in the convention of \cite{Kim:2024svw}.
Counting the number of ways to contract the legs, we find
\begin{align}
    \frac{1}{12} \times 3 = \frac{1}{4} \,.
\end{align}
The matching of the coefficients works much less trivially for the second example, $\tilde{\chi}_{(2)3}$.  
There are many trees of $\chi_{(5)}$ participating in the game. 
The relevant equality is 
\begin{align}
    \frac{1}{5!} ( 9 + 12 - 1 + 1 + 1 + 8) = \frac{1}{4} \,.
\end{align}
For the origin of the integers in the numerator, see the $(5,Y)$ part of Figure~1 in \cite{Kim:2024svw}.

As a final remark, for a Kerr-Schild metric background, 
the relation between $h_{\mu\nu}$ and $\tilde{h}^{\mu\nu}$ becomes linear, 
so that the WQFT and the Magnus expansion agrees term by term. 


\section{External fields} \label{sec:field}

The classical eikonal treats particles and fields on an equal footing. 
In a sense to be elaborated, we can use the results of section~\ref{sec:EM-background} and \ref{sec:Grav-background} with the roles of particle and field exchanged. The eikonal can be used to compute the ``impulse" (radiation) of the fields due to the particles. 
The photon field has no self-interaction, so the coupling to charged particles is the only source.
For the graviton field, we should also consider the self-interaction vertices from the Einstein-Hilbert action as discussed in \cite{Kim:2025hpn}.

\paragraph{0.5PL}
Recall the formula \eqref{chi1-EM} for the leading eikonal in an EM background: 
\begin{align}
    \chi_{(0.5)} = q \int v^\mu A_\mu(x+v\tau)\, d\tau \,, 
    \label{chi-0.5PL}
\end{align}
which has been renamed ``0.5PL eikonal". 
Given a background field configuration, it generates the leading impulse for the charged particle.
On the flip side, we can regard $\chi_{(0.5)}$ as the generator of the field produced by the charged particle: 
\begin{align}
    \Delta_{(0.5)} F_{\mu\nu}(y) = \{ \chi_{(0.5)} , F_{\mu\nu}(y)\} \,.
\end{align}
Using \eqref{FA-bracket}, and noting that $D(x-y) = G(y-x)$ when the time ordering $y^0 > x^0$ is maintained,  we obtain
\begin{align}
\begin{split}
  \{ \chi_{(0.5)} ,F^{\mu\nu}(y) \} &= q \int d\tau \,v^\rho \{A_\rho(\bar{x}),F^{\mu\nu}(y)\}
\\
&= q \int d\tau  \left[ \partial^\mu G(y-\bar{x}) v^\nu -\partial^\nu G(y-\bar{x}) v^\mu\right] \,,   
\end{split}
\label{0.5PL-final}
\end{align}
It agrees with \eqref{F-from-current} with the current due to a static charged particle. 
Thus even the Coulomb potential can be understood from the eikonal point of view. 

\paragraph{1PL - Compton scattering} 

The most prominent observable at 1PL is the momentum impulse of the particles in a binary system. 
But, since it has been discussed extensively in the literature up to 5PL \cite{Bern:2021xze,Wang:2022ntx,Bern:2023ccb}, 
we choose to focus on the Compton scattering, the only 1PL process involving external photons.

In essence, we need to reinterpret the eikonal from section~\ref{sec:EM-background}, 
\begin{align}
   \chi_{(1)\mathrm{Compton}} = \frac{q^2}{2m}  \int  R(\tau_{12}) E_{\mu}(\tau_1) E^{\mu}(\tau_2) d\tau_1 d \tau_2 \,, 
   \quad (\tau_{12} = \tau_1 - \tau_2 )
   \label{chi-Compton}
\end{align}
as the generator of the Compton scattering.
The radiation-eikonal relation is
\begin{align}
    \Delta_{(1)} F^{\mu\nu}(y) = \{ \chi_{(1)} , F^{\mu\nu}(y) \} + \frac{1}{2} \{ \chi_{(0.5)} , \{ \chi_{(0.5)} ,F^{\mu\nu}(y) \} \} \,.
    \label{1PL-radcom}
\end{align}

Let us first compute the LHS from the EOM. 
We should apply the current-to-field formula \eqref{F-from-current} 
to the current accounting for the acceleration of the particle, 
\begin{align}
    \delta J^\nu(z) = q \int\left[ \delta^{4}(z-\bar{x}_1)\delta \dot{x}^\nu(\tau_1) - \partial_\rho \delta^4(z-\bar{x}_1)\delta x^\rho(\tau_1)  v^\nu \right] d\tau_1 \,.
\end{align}
After the $z$-integration, we get 
\begin{align}
\begin{split}
      \Delta_{(1)} F^{\mu\nu}(y)  &= q \int \left[ \partial^\mu G(y-\bar{x}_1)\delta \dot x^\nu (\tau_1) - \partial_\rho \partial^\mu G(y-\bar{x}_1) \delta x^\rho(\tau_1) v^\nu \right] d\tau_1 
      \\
      &\qquad - (\mu\leftrightarrow \nu)\,.
\end{split}
\end{align}
Doing an IBP with $\delta \dot{x}(\tau_1)$, we get 
\begin{align}
    \Delta_{(1)} F^{\mu\nu}(y)  &= q \int \partial_\rho \partial^\mu G(y-\bar{x}_1) \left[ \delta x^\nu(\tau_1) v^\rho-  \delta x^\rho(\tau_1) v^\nu \right] d\tau_1 - (\mu\leftrightarrow \nu)\,.
\end{align}
Importing the known expression for $\delta x(\tau_1)$, we obtain 
\begin{align}
   \Delta_{(1)} F^{\mu\nu}(y)  &= \frac{q^2}{m} \int R_{12}  \partial_\rho \partial^\mu G(y-\bar{x}_1)  \left[  E^\nu(\bar{x}_2) v^\rho - E^\rho(\bar{x}_2) v^\nu \right] d\tau_1 d\tau_2- (\mu\leftrightarrow \nu) \,.
   \label{Compton-EOM}
\end{align}

\begin{figure}[htbp]
    \centering
    \includegraphics[width=0.5\linewidth]{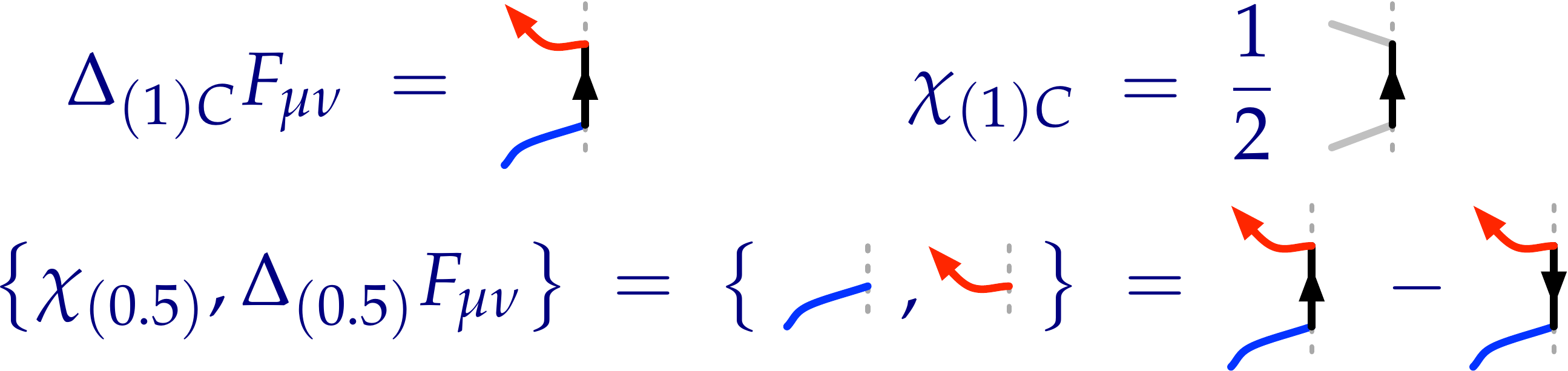}
    \caption{The eikonal for the Compton scattering.}
    \label{fig:1PL-Compton}
\end{figure}

On the RHS of \eqref{1PL-radcom}, consider the iteration term first, 
\begin{align}
\begin{split}
    \frac{1}{2} \{ \chi_{(0.5)} , \{ \chi_{(0.5)} ,F^{\mu\nu}(y) \} \} = 
    \frac{1}{2} \{ \chi_{(0.5)} , \Delta_{(0.5)} F^{\mu\nu}(y)  \} \,.
    \label{1PL-rad-2nd}
\end{split}
\end{align}
Given the formula \eqref{0.5PL-final} for $\Delta_{(0.5)} F^{\mu\nu}(x)$, 
the computation involves terms like 
\begin{align}
\begin{split}
       m \{ A_\beta(\bar{x}_2)v^\beta , \partial^\mu G(y-\bar{x}_1)v^\nu \} &= \left[ - \tau_{12} \partial_\rho A_\beta(\bar{x}_2)v^\beta + A_\rho(\bar{x}_2)\right] \partial^\rho \partial^\mu G(y-\bar{x}_1) v^\nu 
       \\
       &\qquad + \partial^\nu A_\beta(\bar{x}_2) v^\beta \partial^\mu G(y-\bar{x}_1) \,.
\end{split}
\end{align}
The gauge invariance is obscured, but as we experienced in section~\ref{sec:EM-background}, the non-invariant terms become total derivatives. Up to IBP relations, we can show that  
\begin{align}
\begin{split}
       &m \{ A_\beta(\bar{x}_2)v^\beta , \partial^\mu G(y-\bar{x}_1)v^\nu \} 
       \\
       &\quad \approx  - \tau_{12} E^{\rho}(\bar{x}_2)  \partial_\rho \partial^\mu G(y-\bar{x}_1) v^\nu  + E^\nu(\bar{x}_2)  \partial^\mu G(y-\bar{x}_1) \,.
\end{split}
\end{align}
We can do a further IBP using $d(\tau_{12})/d\tau_1 = 1$ and get 
\begin{align}
\begin{split}
       m \{ A_\beta(\bar{x}_2)v^\beta , \partial^\mu G(y-\bar{x}_1)v^\nu \} 
       \approx  \tau_{12}\left[ -  E^{\rho}(\bar{x}_2)   v^\nu  +  E^\nu(\bar{x}_2)   v^\rho \right] \partial_\rho \partial^\mu G(y-\bar{x}_1)\,.
\end{split}
\end{align}
So, the iteration term becomes
\begin{align}
\begin{split}
    &\frac{1}{2} \{ \chi_{(0.5)} , \Delta_{(0.5)} F^{\mu\nu}(y)  \} 
    \\
    &= \frac{q^2}{2m} \int d\tau_1d\tau_2 (R_{12}-R_{12}) \left[ -  E^{\rho}(\bar{x}_2)   v^\nu  +  E^\nu(\bar{x}_2)   v^\rho \right] \partial_\rho \partial^\mu G(y-\bar{x}_1) - (\mu \leftrightarrow \nu) \,.
\end{split}
\label{Compton-iteration}
\end{align}
The first term in the RHS of \eqref{1PL-radcom} with 
the Compton-eikonal \eqref{chi-Compton} gives 
\begin{align}
\begin{split}
    &\{\chi_{(1)\mathrm{C}} , F^{\mu\nu}(y) \}
    \\
    &=\frac{q^2}{2m}\int d\tau_1d\tau_2 (R_{12}+R_{21})E_\rho(\bar{x}_2) v_\sigma  \{F^{\rho \sigma}(\bar{x}_1),F^{\mu \nu}(y)\} 
    \\
    &=\frac{q^2}{2m}\int d\tau_1d\tau_2 (R_{12}+R_{21}) \left[ - E^\rho(\bar{x}_2) v^\nu + E^\nu(\bar{x}_2) v^\rho \right] \partial_\rho \partial^\mu G(y-\bar{x}_1) - (\mu \leftrightarrow \nu) \,,
\end{split}
\label{Compton-chi2}
\end{align}
where we used  \eqref{FF-bracket}. 
Clearly, the sum of \eqref{Compton-iteration} and \eqref{Compton-chi2} 
agrees with \eqref{Compton-EOM}.
The key elements of the Compton scattering problem are summarized pictorially in Figure~\ref{fig:1PL-Compton}. 

\paragraph{Momentum conservation}

One important consistency check in a radiative process is the total momentum conservation, 
\begin{align}
    \Delta p^\mu_\mathrm{particle} +  \Delta p^\mu_\mathrm{field} = 0 \,.
\end{align}
If the photon field has a static component, the electron will get kicked at $\mathcal{O}(q)\sim$ 0.5PL. But, if the incoming photon field forms a wave packet around a fixed non-zero frequency, 
the 0.5PL impulse will average out and the real impulse will begin at 1PL. 

The external fields satisfy the source free Maxwell's equation, $\partial^\mu F_{\mu\nu} = 0$, 
which implies that $F_{\mu\nu}$ and its derivatives $\partial_\rho F_{\mu\nu}$, $\partial_\rho\partial_\sigma F_{\mu\nu}$  all satisfy the massless Klein-Gordon equation: $\partial^2 F_{\mu\nu} = 0$, etc.  
In what follows, an identity valid for any such solution will be useful:
\begin{align}
    \begin{split}
        \phi(x)=-\int d^3\vec{y}\left(G(y-x)\partial_0 \phi(y) - \partial_0G(y-x) \phi(y)\right)\,.
    \end{split}
    \label{Green-identity}
\end{align}
To prove it, consider the vector field 
\begin{align}
    K_\mu =  G(y-x) \partial_\mu \phi(y) -  \partial_\mu G(y-x)\phi(y) \,.
\end{align}
Its divergence is 
\begin{align}
    \partial^\mu K_\mu = G(y-x) \partial^\mu \partial_\mu\phi(y) -  \partial^\mu \partial_\mu G(y-x)\phi(y) = \delta(y-x) \phi(y) \,.
\end{align}
Applying Stokes' theorem to the integral $\int d^4 z \,\partial^\mu K_\mu(z,x)$ between $z^0 = x^0- \epsilon $ and $z^0 = y^0$, and recalling the fact that $G(z-x)$ vanishes for $z^0 < x^0$, we arrive at \eqref{Green-identity}. 

We begin with the familiar expression for the photon momentum, 
\begin{align}
    p^\mu_\mathrm{field} = \int d^3 \vec{y}\,(-T^\mu{}_\nu v^\nu) = - \int d^3 \vec{y}\left( F^{\mu\rho} F_{\nu\rho}  - \frac{1}{4} \delta^\mu{}_\nu F^{\rho\sigma} F_{\rho\sigma}\right) v^\nu\,.
    \label{field-momentum}
\end{align}
To compute the 1PL impulse of the photon, we examine the integrand, 
\begin{align}
  \Delta \mathcal{P}^\mu \equiv -\Delta T^\mu{}_\nu v^\nu = -\left( \Delta F^{\mu\rho} F_{\nu\rho} +  F^{\mu\rho} \Delta F_{\nu\rho} - \frac{1}{2} \delta^\mu{}_\nu \Delta F^{\rho\sigma} F_{\rho\sigma}\right) v^\nu \,.
  \label{T-mu-1PL}
\end{align}
In view of the general impulse-eikonal relation, this integrand excludes the $\{ \chi_{0.5} , F\}^2$ contribution. But, 
as we noted in \eqref{0.5PL-final}, the only 0.5PL contribution is the Coulomb potential 
and the energy carried by the Coulomb field is irrelevant for the Compton scattering. 

Importing the 1PL field from \eqref{Compton-EOM}, 
\begin{align}
   \Delta F^{\mu\rho}(y)  &= \frac{q^2}{m} \int R_{12}  \partial_\sigma \partial^\mu G(y-\bar{x}_1)  \left[  E^\rho(\bar{x}_2) v^\sigma - E^\sigma(\bar{x}_2) v^\rho \right] d\tau_1 d\tau_2- (\mu\leftrightarrow \rho) \,, 
   \label{Compton-EOM-copy}
\end{align}
and plugging it into \eqref{T-mu-1PL}, after partial cancellations, we find 
\begin{align}
\begin{split}
    \Delta \mathcal{P}^\mu &= \frac{q^2}{m} \int R_{12} \,\mathcal{I}^\mu d\tau_1 d\tau_2 \,,
     \\
     \mathcal{I}^\mu &=  [(v\cdot \partial) (\partial^\mu G)] E_\rho(\tau_2) E^\rho(y) -    [(v\cdot \partial) \partial_\rho G]\left[  E^\mu(\tau_2) E^\rho(y) -  E^\rho(\tau_2) E^\mu(y) \right] 
     \\
     &\qquad - [(v\cdot \partial)^2 G] E_\rho(\tau_2) F^{\mu\rho}(y) + [\partial_\nu \partial_\rho G] E^\nu(\tau_2) F^{\mu\rho}(y) 
     \\
     &\qquad + v^\mu [(v\cdot \partial)\partial_\rho G] E_\nu(\tau_2) F^{\rho\nu}(y) \,.
\end{split}
\end{align}
The derivatives are in $y^\mu$ and act on $G = G(y-\bar{x}_1)$ only. By abuse of notation, we set $E_\rho(\tau_2) = E_\rho(\bar{x}_2) = E_\rho(\bar{x}(\tau_2))$. 

In the frame where $v^\mu=(1,0,0,0)$, the ``energy" component gives 
\begin{align}
\begin{split}
      \Delta \mathcal{P}^0
      &= \frac{q^2}{m} \int R_{12} \left( -[\partial_0^2 G] E_i(y) + [\partial_i \partial_j G] E^j(y)  + [\partial_0 \partial_j G]  F^j{}_i(y) \right) E^i(\tau_2) \,.
\end{split}
\end{align}
Integrating it over the spatial slice, and applying IBPs and Maxwell's equations in vacuum, we get 
\begin{align}
\begin{split}
      \Delta p^0_\mathrm{field} &= \int d^3\vec{y} \, \Delta \mathcal{P}^0 
      = -\frac{q^2}{m} \int   R_{12} \int d^3\vec{y} \left( [\partial_0^2 G]  E_i(y) - [\partial_0 G] \partial_0 E_i(y) \right) E^i(\tau_2)\,.
\end{split}
\end{align}
Applying further IBPs and applying \eqref{Green-identity} to $\partial_0 E_i$, we obtain 
\begin{align}
    \Delta p^0_\mathrm{field} &= -\frac{q^2}{m} \int R_{12} (\partial^0 E^\nu(\tau_1)) E_\nu(\tau_2) d\tau_1 d\tau_2 \,.
\end{align}
For the ``spatial momentum" components, we find 
\begin{align}
\begin{split}
      \Delta \mathcal{P}^i &= \frac{q^2}{m} \int R_{12} \int d^3\vec{y} \,\left( [\partial^i \partial_0 G] E^j(y) - [\partial_0^2 G] F^{ij}(y) + [\partial_j \partial_k G] F^{ik}(y) \right) E_j(\tau_2) 
      \\
      &\quad - \frac{q^2}{m} \int R_{12} \int d^3\vec{y} \,[\partial_0 \partial_j G] E^j(y) E^i(\tau_2) \,.
\end{split}
\end{align}
Integrating it over the spatial slice, and applying IBPs, we get \
\begin{align}
\begin{split}
      &\Delta p^i_\mathrm{field} = \int d^3\vec{y} \, \Delta \mathcal{P}^i
      \\
      &\quad= \frac{q^2}{m} \int   R_{12} \int d^3\vec{y} \left( [\partial^i \partial_0 G] E^j(y) - [\partial_0^2 G]  F^{ij} (y) + [\partial^j \partial_k G] F^{ik}(y)\right) E_j(\tau_2)\,.
\end{split}
\end{align}
Applying further IBPs and applying \eqref{Green-identity} to $\partial_i E_j$, we obtain 
\begin{align}
   \Delta p^i_\mathrm{field} &= -\frac{q^2}{m} \int R_{12} (\partial^i E^\nu(\tau_1)) E_\nu(\tau_2) d\tau_1 d\tau_2 \,.
\end{align}
Combining all components, we conclude that 
\begin{align}
     \Delta p^\mu_\mathrm{field} &= -\frac{q^2}{m} \int R_{12} (\partial^\mu E^\nu(\tau_1)) E_\nu(\tau_2) d\tau_1 d\tau_2 \,.
\end{align}
The total momentum is conserved as expected. 

Both for the energy component and the spatial momentum components, the IBP involving $\partial_0^2 G$ deserves attention. 
We use the defining equation $(-\partial^\mu\partial_\mu)G(y-x) = \delta(y-x)$ to replace $\partial_0^2 G(y-x)$ with $\partial_i^2 G(y-x) + \delta(y-x)$. We can apply IBPs to $\partial_i^2 G(y-x)$ inside the $d^3\vec{y}$ integral. 
The delta function leads to the incoming photon field measured at the location of the charged particle at the future infinity, which dies off quickly.

We presented the computation of the impulse of the field in detail, an instructive exercise to show the usefulness of the classical eikonal for fields. But momentum conservation is a direct consequence of the translation invariance, 
so as long as the eikonal is translation invariant order by order, momentum conservation holds automatically. 
Concretely, the momentum as the translation generator for the EM field is encoded in the bracket, 
\begin{align}
    \{ F_{\mu\nu}(x), p^\rho_\mathrm{field} \} = - \partial^\rho F_{\mu\nu}(x) \,.
\end{align}
It is straightforward to verify that \eqref{field-momentum} satisfies the relation. The proof involves 
manipulating the derivatives of $G(y-x)$ as in the paragraphs above, so we shall not repeat it here.

\paragraph{1.5PL} 

\begin{figure}[htbp]
    \centering
    \includegraphics[width=0.48\linewidth]{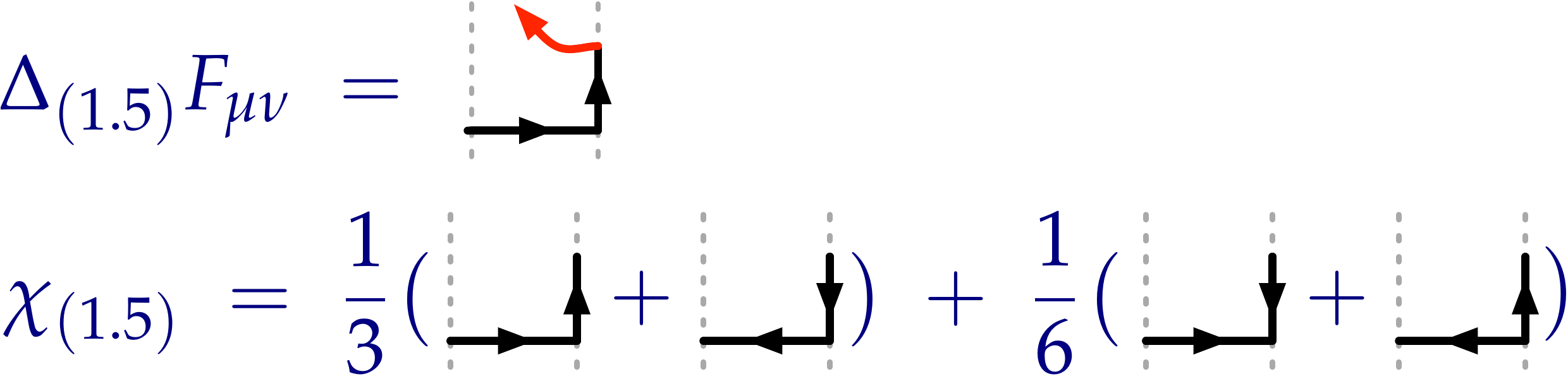}
    \caption{The 1.5PL radiation.}
    \label{fig:1.5PL-radiation}
\end{figure}

For a binary system of massive particles, the radiation begins at 1.5PL. 
When one is much heavier than the other, the 1.5PL radiation mainly comes from the curved trajectory of the light particle in the Coulomb background of the heavy particle. The impulse-eikonal relation involves 
\begin{align}
\begin{split}
    \Delta_{(1.5)} F_{\mu\nu} &= \{ \chi_{(1.5)} , F_{\mu\nu}\} + \frac{1}{2} \{ \chi_{(1)} , \{ \chi_{(0.5)} , F_{\mu\nu}\}\}  
    \\
    &\qquad + \frac{1}{2} \{ \chi_{(0.5)} , \{ \chi_{(1)} , F_{\mu\nu}\}\} + \frac{1}{6} \{ \chi_{(0.5)} , \{ \chi_{(0.5)} , \{ \chi_{(0.5)} , F_{\mu\nu}\}\}\} 
    \\
    &= \{ \chi_{(1.5)} , F_{\mu\nu}\} + \frac{1}{2} \{ \chi_{(1)} , \Delta_{(0.5)}  F_{\mu\nu}\} 
    \\
    &\qquad + \frac{1}{2} \{ \chi_{(0.5)} , \Delta_{(1)}  F_{\mu\nu}\} - \frac{1}{12} \{ \chi_{(0.5)} , \{ \chi_{(0.5)} , \Delta_{(0.5)}  F_{\mu\nu}\}\} \,.
    \label{1.5PL-iteration-bf}
\end{split}
\end{align}
Using the previous results from lower orders, we again confirm the expectation from the Magnus expansion. The key elements are summarized in Figure~\ref{fig:1.5PL-radiation}.

\acknowledgments

We are grateful to Joon-Hwi Kim, Jung-Wook Kim and Tabasum Rahnuma for discussions. 
We thank Tianheng Wang for collaboration at early stages of this project, 
and the participants of the Amplitudes 2025 conference [APCTP-2025-C04] for fruitful discussions.
This work is supported in part by the
National Research Foundation of Korea (NRF) grants, NRF-2023-K2A9A1A0609593811
and NRF RS-2024-00351197.
%


\newpage
\bibliographystyle{JHEP}
\bibliography{biblio}

\end{document}